\documentclass[conference]{IEEEtran}

\usepackage[numbers]{natbib}
\usepackage{amsmath,amsfonts}
\usepackage{figsize}
\usepackage{textcomp}
\usepackage{glossaries}
\usepackage{bm}
\usepackage{xcolor}
\usepackage{svg}
\usepackage{algorithm}
\usepackage{algpseudocode}
\usepackage{caption} 
\usepackage{tabu}
\usepackage{multirow}
\usepackage{makecell}
\usepackage{array}
\usepackage{hyperref}
\usepackage{todonotes}
\usepackage{regexpatch}
\usepackage{xspace}
\usepackage{setspace}
\usepackage{enumitem}
\usepackage{soul}

\makeatletter
\xpatchcmd{\@todo}{\setkeys{todonotes}{#1}}{\setkeys{todonotes}{inline,backgroundcolor=yellow,#1}}{}{}
\makeatother

\newcommand{\name}{NetSentry\xspace}

\newcolumntype{?}{!{\vrule width 1pt}}

\algnewcommand{\Inputs}[1]{%
  \State \textbf{Inputs:}
  \Statex \hspace*{\algorithmicindent}\parbox[t]{.8\linewidth}{\raggedright #1}\vspace{3pt}
}
\algnewcommand{\Initialize}[1]{%
  \State \textbf{Initialisation:}
  \Statex \hspace*{\algorithmicindent}\parbox[t]{.8\linewidth}{\raggedright #1}\vspace{3pt}
}

\def\BibTeX{{\rm B\kern-.05em{\sc i\kern-.025em b}\kern-.08em
    T\kern-.1667em\lower.7ex\hbox{E}\kern-.125emX}}

\newacronym{IDS}{IDS}{Intrusion Detection Systems}
\newacronym{NIDS}{NIDS}{Network Intrusion Detection Systems}
\newacronym{NID}{NID}{Network Intrusion Detection}
\newacronym{HIDS}{HIDS}{Host-based Intrusion Detection Systems}
\newacronym{DDoS}{DDoS}{Distributed Denial of Service}
\newacronym{DoS}{DoS}{Denial of Service}
\newacronym{DL}{DL}{Deep Learning}
\newacronym{IoT}{IoT}{Internet of Things}
\newacronym{CSRF}{CSRF}{Cross-Site Request Forgery}
\newacronym{LSTM}{LSTM}{Long Short-Term Memory}
\newacronym{Bi-LSTM}{Bi-LSTM}{Bidirectional LSTM}
\newacronym{ConvLSTM}{ConvLSTM}{Convolutional LSTM}
\newacronym{Bi-ALSTM}{Bi-ALSTM}{Bidirectional Asymmetric LSTM}
\newacronym{RNN}{RNN}{Recurrent Neural Networks}
\newacronym{MLP}{MLP}{Multilayer Perceptron}
\newacronym{CNN}{CNN}{Convolutional Neural Networks}
\newacronym{IAT}{IAT}{Inter-arrival Time}
\newacronym{CIC}{CIC}{Canadian Institute for Cybersecurity}
\newacronym{CSE}{CSE}{Communications Security Establishment}
\newacronym{OC-NN}{OC-NN}{One Class Neural Nets}
\newacronym{DAGMM}{DAGMM}{Deep Autoencoding Gaussian Mixture Model}
\newacronym{TP}{TP}{True Positives}
\newacronym{FP}{FP}{False Positives}
\newacronym{TN}{TN}{True Negatives}
\newacronym{FN}{FN}{False Negatives}
\newacronym{ML}{ML}{Machine Learning}
\newacronym{SGD}{SGD}{Stochastic Gradient Descent}
\newacronym{XSS}{XSS}{Cross-Site Scripting}
\newacronym{SMB}{SMB}{Server Message Block}
\newacronym{LAN}{LAN}{Local Area Network}
\newacronym{WAN}{WAN}{Wide Area Network}
\newacronym{PDF}{PDF}{Probabilistic Density Function}
\newacronym{MACs}{MACs}{Multiply-Accumulate Operation Counts}
\newacronym{LRCN}{LRCN}{Long-term Recurrent Convolutional Networks}
\newacronym{BPTT}{BPTT}{Backpropagation Through Time}
\newacronym{NLL}{NLL}{Negative Log Likehood}
\newacronym{SVM}{SVM}{Support Vector Machine}
\newacronym{ROC}{ROC}{Receiver Operating Characteristic}
\newacronym{AUC}{AUC}{Area Under Curve}
\newacronym{IPS}{IPS}{Intrusion Prevention System}
\newacronym{OS}{OS}{Operating System}
\newacronym{REP}{REP}{Relative Error Percentage}
\newacronym{ECDF}{ECDF}{Empirical Cumulative Distribution Function}
\newacronym{FPR}{FPR}{False Positive Rate}
\newacronym{TPR}{TPR}{False Positive Rate}
\newacronym{FNR}{FNR}{False Negative Rate}

\hypersetup{
    colorlinks = false,
    linkbordercolor = {white},
}

\makeatletter
\g@addto@macro{\UrlBreaks}{\UrlOrds}
\makeatother

\begin{document}
\let\WriteBookmarks\relax
\def\floatpagepagefraction{1}
\def\textpagefraction{.001}

\title{\name: A Deep Learning Approach to Detecting Incipient Large-scale Network Attacks}

\author{\IEEEauthorblockN{Haoyu Liu}
\IEEEauthorblockA{\textit{University of Edinburgh} \\
haoyu.liu@ed.ac.uk\vspace*{-8pt}}
\and
\IEEEauthorblockN{Paul Patras}
\IEEEauthorblockA{\textit{University of Edinburgh} \\
paul.patras@ed.ac.uk\vspace*{-8pt}}}

\maketitle

\begin{abstract}
\gls{ML} techniques are increasingly adopted to tackle ever-evolving high-profile network attacks, including \gls{DDoS}, botnet, and ransomware, due to their unique ability to extract complex patterns hidden in data streams. These approaches are however routinely validated with data collected in the same environment, and their performance degrades when deployed in different network topologies and/or applied on previously unseen traffic, as we uncover. This suggests malicious/benign behaviors are largely learned superficially and \gls{ML}-based \gls{NIDS} need revisiting, to be effective in practice.
In this paper we dive into the mechanics of large-scale network attacks, with a view to understanding how to use \gls{ML} for \gls{NID} in a principled way. We reveal that, although cyberattacks vary significantly in terms of payloads, vectors and targets, their \emph{early stages}, which are critical to successful attack outcomes, \emph{share many similarities and exhibit important temporal correlations}. Therefore, we treat \gls{NID} as a time-sensitive task and propose \name, perhaps the first of its kind \gls{NIDS} that builds on \gls{Bi-ALSTM}, an original ensemble of sequential neural models, to detect network threats before they spread. We cross-evaluate \name using two practical datasets, training on one and testing on the other, and demonstrate F1 score gains above 33\% over the state-of-the-art, as well as up to 3$\times$ higher rates of detecting attacks such as \gls{XSS} and web bruteforce. Further, we put forward a novel data augmentation technique that boosts the generalization abilities of a broad range of supervised deep learning algorithms, leading to average F1 score gains above~35\%. 
Lastly, we shed light on the feasibility of deploying \name in operational networks, demonstrating affordable computational overhead and robustness to evasion attacks.

\end{abstract}


\section{Introduction}
The volume of illicit network traffic continues to grown dramatically, with the number of high-profile attacks including \gls{DDoS}, botnet, and ransomware rising by over 45\% annually \cite{botstat}, and the losses incurred expected to exceed 6 trillion US dollars in 2021~\cite{lossstat}. Effective countermeasures to thwart ever-evolving cyber threats are therefore urgently needed. Traditional \glsfirst{NIDS} largely apply finite rules, preset by human experts, to detect anomalies. This approach lacks flexibility and is often prone to subversion~\cite{beforeyouknew}.  \glsfirst{ML} is increasingly used to detect cyber intrusions, due to its ability to discover complex statistical patterns hidden in data streams, which can aid in discriminating anomalies based on feature differences~\cite{buczak2015survey}. 

\gls{ML} is a powerful tool, yet adopting it meaningfully for security purposes is not straightforward. \gls{ML} techniques used in areas including imaging and natural language processing have been directly applied to \gls{NID} (e.g., \cite{lin2018idsgan}), without adequate analysis of their suitability for this task. For instance, reconstruction-based algorithms like autoencoders were originally designed to learn to recreate benign samples that contain similar patterns, e.g., the same object type in images \cite{ xia2015learning}. However, when deployed for intrusion detection, whether an autoencoder is able to reconstruct heterogeneous benign traffic originating from various applications is rarely discussed~\cite{mirsky2018kitsune}.  
%
Secondly, widely-used evaluation methodologies involve training and testing \gls{NID} models on a single dataset, collected in the same controlled environment. 
This makes it difficult to assess if the trained models can truly generalize to previously unseen traffic mixes~\cite{sommer2010}. Moreover, detecting high-volume attacks promptly, before a target system becomes overloaded and unable to thwart malicious traffic with potential to cause severe damage following early system compromise, is difficult. This capability is however critical to the availability and revenue of online businesses~\cite{ba20}. 

In this paper, we address the above challenges and propose \textbf{\name}, a novel \gls{DL}-based \gls{NIDS} that reliably detects a range of malicious traffics with similar patterns, indicative of incipient high-impact network attacks. 
As such, we make the following \textbf{key contributions}:

\begin{enumerate}[itemsep=0pt, label=(\arabic*), topsep=0pt]
    \item We scrutinize several attack chains and identify key temporal inter-relations between illicit traffic occurring in the wild; based on this analysis, we design Bidirectional Asymmetric LSTM (Bi-ALSTM), an original ensemble of sequential neural models that effectively captures the temporal dynamics of malicious traffic and classifies specific threats, including \gls{DoS}, Port Scanning, and Brute Forcing;  
    
    \item  Since not every attack type can be distinguished accurately with limited information available at the network layer, we introduce a novel training technique that relies on feature augmentation and abstract labeling. The feature augmentation scheme improves the heterogeneity of cyber attacks that were collected in a controlled environment, which helps NN models learn a more robust decision boundary. Abstract labeling, on the other hand, prevents overfitting by grouping similar types of attacks into one class;
    
    
    \item We train our Bi-ALSTM on a large dataset published by the Canadian Institute for Cybersecurity,  we cross-evaluate our approach with a previously unseen dataset collected in a different network topology, and we compare its performance against that of state-of-the art benchmarks. Results demonstrate Bi-ALSTM  outperforms existing approaches by at least 33\% in terms of F1 score; 
    
    \item We discuss practical aspects of deploying \name in real-life, including computational overhead and 
    robustness to a range of evasion attacks.
\end{enumerate}

To our knowledge, \name is perhaps the first principled \gls{DL}-based \gls{NIDS} that tackles cyberthreats 
by focusing on the early stages that are essential to the success of large-scale and  high-impact attacks such as botnet and ransomware.

\section{Threat Model \& Anatomy of Attacks}
\label{sec:threat_model}
We start from the key observation that in practice traffic flows shall not be considered in isolation, either as benign or malicious. There exist important temporal correlations among different cyber attacks, especially those with high-impact, which rarely occur independently. For instance, assume that an adversary has zero knowledge of a potentially vulnerable target. Conducting a successful webshell injection attack has at least two pre-requisites:
\emph{(i)} port scanning against the target, so as to uncover that it runs a web serve; and \emph{(ii)} web API enumerating, to verify if file upload is allowed. 
That is, the attacker must follow a certain sequence of actions (each an attack itself), which would create distinct network traces at various stages. We remark that essential correlations among different kinds of network attacks have not been explored before, but are potentially useful to design a reliable \gls{NIDS}. 

Hence, we decompose network attacks from the perspective of an active adversary, and summarize them into different attack chains. These typically start with gathering information about a target and conclude when a specific technical goal was achieved. We consider three key attack chains, namely botnet, web intrusion, and ransomware, revealing that they are supported by a similar methodology. 
While our attack~chain view may appear on the surface related to earlier Botnet infection modeling, where attack stages are fingerprinted~\cite{gu2007bothunter}, our modeling approach and subsequent NIDS design are fundamentally different. This is because \textit{the attack chains we consider aim to reveal the common stages shared by different large-scale cyber attacks}, so as to impede a specific range of cyber intrusions by interrupting any of these early stages. 

\subsection{Attack Chain Analysis}

\begin{figure}[t]
    \centering
    \includegraphics[width=\columnwidth]{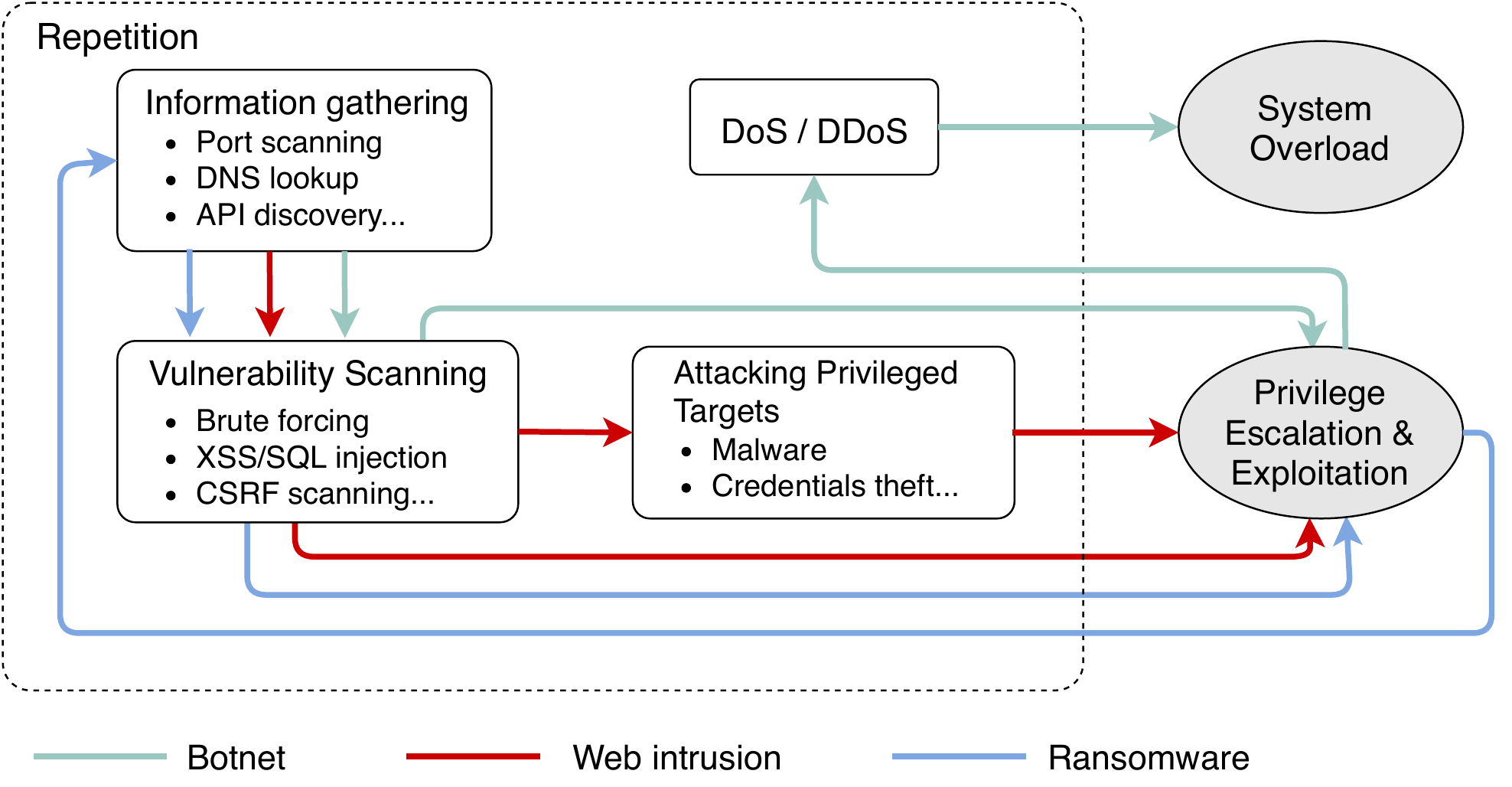}
    \caption{Attack chains employed by large-scale high-impact threats, e.g., botnets, web intrusion, and ransomware. Observe that similar steps are repeated by all, which \textbf{\name} exploits for detection.}
    \label{fig:attack_chain}
\vspace*{-1em}
\end{figure}

We particularly focus on two network attack goals, which bring severe damage to target systems. The first is to obtain system privileges temporarily or permanently, by exploiting various security flaws. The second is to overload the system by occupying all its resources. Instead of looking at each attack type individually, we investigate what processes, i.e., attack chains, an adversary must follow to achieve any of these goals, when having zero knowledge about a target. We consider three unique attack chains that are specific to botnet, web intrusion, and respectively ransomware, as shown
in~Figure~\ref{fig:attack_chain}.\footnote{We note that certain sophisticated attacks may have longer attack chains that expand to application level~\cite{ifflander2019hands}. However, by detecting their early stages, their later exploitation actions can be prevented.} 

\textbf{Botnet/Mirai.} Botnets are collections of Internet-accessible devices hijacked by an attacker, and are usually employed to carry out large-scale, high-impact \gls{DDoS} attacks. Mirai is one of the most notorious instances in recent years, with $\sim$600,000 devices infected at its peak 
\cite{antonakakis2017understanding}. Subsequent Mirai variants expand the attack surface to SSH, HTTPS, FTP, etc., but inherit the methodology of the canonical version.

Mirai follows a chain-like methodology that entails \textit{information gathering} $\rightarrow$ \textit{vulnerability scanning} $\rightarrow$ \textit{privilege escalation} $\rightarrow$ \textit{\gls{DDoS}}.  Specifically, (1) TCP SYN packets are sent towards the entire IPv4 address space, on ports 23 and 2323 (Telnet); (2) after identifying potential victims, Mirai attempts to bruteforce the victims' credentials 
using a dictionary -- this process is deemed as vulnerability scanning; (3) upon successful login, a victim's IP and 
credentials are forwarded to a report server that infects the victim with malicious code; (4) newly infected devices become members of the botnet, and either participate in victim discovery or \gls{DDoS} attacks \cite{antonakakis2017understanding}.

\begin{figure*}[t]
    \centering
    \includegraphics[width=0.96\linewidth]{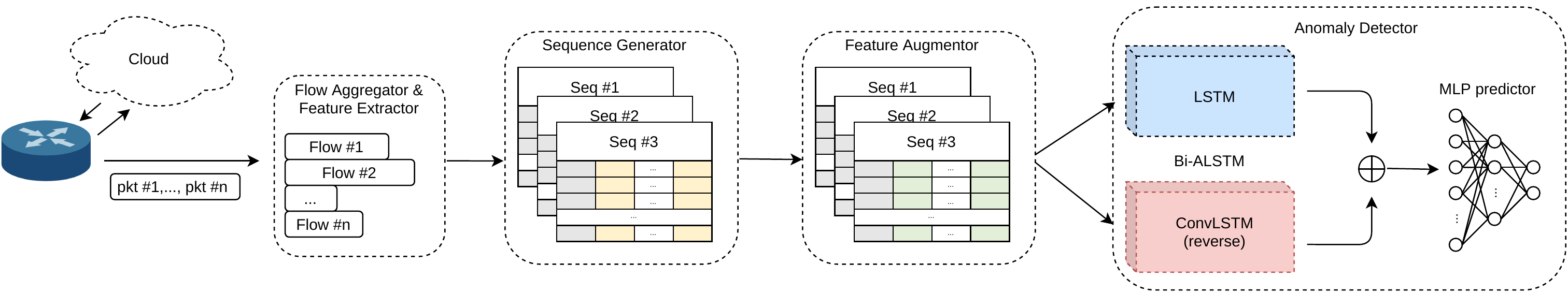}
    \caption{\name architecture. Feature Augmentor only applied during training. $\oplus$ is fusion operation detailed in Eq.~\ref{eqn:combine}.}
    \label{fig:arch}
\vspace*{-1em}
\end{figure*}

\textbf{Web intrusion.} Web applications 
integrate a technology stack that includes storage, web engines, \glspl{OS}, and communications. Hence, various vulnerabilities are often exploited.
Web intrusion can be jointly modeled with the Intrusion Kill Chain \cite{hutchins2011intelligence} and the OWASP Web Penetration Guideline \cite{owasp}, where the former outlines a general intrusion process, while the latter provides specific attack vectors. 

The attack process entails \textit{information gathering} $\rightarrow$ \textit{vulnerability scanning} $\rightarrow$ \textit{attacking privileged targets} $\rightarrow$ \textit{exploitation}; or \textit{information gathering} $\rightarrow$ \textit{vulnerability scanning} $\rightarrow$ \textit{privilege escalation}. 
%
Web intrusion resembles botnet and ransomware in terms of vulnerability discovering approach, but differs at the later stage, since exploitation is always target-specific, e.g. a \gls{XSS} attack targets web app users, while SQL injection targets Web APIs.
Our focus is on the early stage 
when the attacker tries to breach a trusted boundary, 
since later steps occur at system level and are invisible to a \gls{NIDS}.

\textbf{Ransomware/WannaCry.}
Ransomware is a relatively new type of threat that blocks user access to their private data until a ransom is paid to the attacker. 
For instance, by exploiting EternalBlue~\cite{eternalblue}, WannaCry gains system access via the \gls{SMB} protocol on Windows systems, encrypts user data, and spreads itself to other hosts~\cite{kao2018dynamic}. The attack chain of WannaCry follows a loop consisting of: \textit{information gathering} $\rightarrow$ \textit{vulnerability scanning} $\rightarrow$ \textit{privilege escalation \& exploitation} $\rightarrow$ \textit{information gathering}.

We focus on the procedure ransomware uses to discover and control new victims (instead of the file encryption applied), as the attack is conducted via the network. Wanna\-Cry employs repeated TCP scanning on port 445 (serving the \gls{SMB} protocol)
~\cite{kao2018dynamic}. Targets are further fingerprinted and remote access is achieved by injecting code via a crafted packet, which would be mishandled by SMBv1. WannaCry then encrypts the data on the victim machine and  discovers other vulnerable targets.

\section{\name Design}
\label{sec:system_design}
In what follows we present \textbf{\name}, an original \gls{NIDS} design that harnesses the unique feature extraction capabilities of recurrent neural models, to detect large-scale, high-impact attacks. \name builds on our observation of correlations between malicious traffics, and handles \gls{NID} as a time-sensitive task, leveraging an ensembling structure to capture richer contexts and detect intrusions with high efficacy. 

\subsection{Attack Detection Strategy}
\label{sec:defensive_scheme}
Our attack chains analysis revealed that information gathering, vulnerability scanning and \gls{DoS} are applied across various types of attacks and share the same semantic. Latent network attacks, such as malware downloading, code injection, \gls{CSRF}, and other zero-day attacks, which can obtain system privileges and proceed to exploitation, always follow massive vulnerability scanning, since this is the most efficient way to discover weak entry points. A common argument is that zero-day attacks are heterogeneous, which poses difficulties to any detection logic. 
However, we argue that \emph{as long as automated activities can be recognized in time, the subsequent zero-day attacks can be blocked,} in order to minimize the chances that attackers may uncover weaknesses and compromise a system. As such, we keep the scope of our detection targets narrow, yet well-directed, as suggested in \cite{sommer2010}. With this in mind, we design \name, a \gls{NIDS} that effectively tackles cyber intrusion by detecting risks at an \emph{early stage}. This also applies to tactics that deviate from the standard attack chains described, as long as they incorporate any common stages to achieve the same end goal.

We maintain that \gls{NID}, especially of automated attacks, should be treated as a time-sensitive task. Here we consider `time-sensitive' those network intrusions exhibiting temporal correlations among \emph{consecutive traffic flows}, which could potentially exert substantial impact on the decision-making process. This is because a single traffic flow, whose features are extracted as a datum, may not fully reflect the intention of the communications. A straightforward example is a TCP flow encapsulating a complete HTTP request. Assume the flow is terminated quickly after the server responds. Without looking at previous and subsequent traffic, it is impossible to assert whether the flow was initiated by a legitimate user or by \gls{DoS} tools. Conversely, if we observe a series of statistically similar communications between a pair of hosts, the confidence of classifying them as malicious becomes higher. Network attacks generated with the same tool usually serve the same purpose. Although they would encapsulate different payloads in consecutive flows for obfuscation or fuzzing purposes, this difference is invisible to a \gls{NIDS} that only has access to protocol headers and timing information. Thus, we leverage sequential neural models in \name to learn such similarity of successive flows generated with automated tools. 

\subsection{System Architecture}
\name is a \gls{NIDS} that examines the statistical features of network flows and detects illicit traffic via an ensemble of sequential neural models. A traffic flow is built by grouping packets according to a five-tuple (Src IP, Dst IP, Src Port, Dst Port, Protocol). 
Recall 
that automated tools tend to initiate multiple almost identical flows towards targets during a short period of time. This means that the statistical features of malicious flows share a large degree of similarity. Thus, monitoring the similarities and discrepancies of consecutive flows between pairs of hosts plays an essential role in recognizing anomalies. To learn relevant temporal correlations of the traffic flows and to differentiate malicious patterns, \name incorporates 4 key building blocks, as shown in Figure~\ref{fig:arch}, namely:
\begin{itemize}[itemsep=0pt,topsep=0pt]
    \item Flow Aggregator \& Feature Extractor: groups packets into flows and extracts associated statistical features; 
    \item Sequence Generator: groups flows originating from the same pair of hosts into fixed-length sequences, to be fed as inputs to anomaly detection logic;
    \item Feature Augmentor: increases the variability of a fraction of malicious traffic features that are non-essential in anomaly detection, but if left unchanged may increase the risk of model becoming trapped in local optima;
    \item Anomaly Detector (Bi-ALSTM): an ensemble of two asymmetric \gls{LSTM}-based neural networks operated bidirectionally, taking flow sequences as input to detect malicious traffic.
\end{itemize}
Next, we explain the inner workings of each component, then detail our bidirectional sequential neural model in \S\ref{sec:ensemble}.

\begin{table*}[]
\centering
\small
\bgroup
\def\arraystretch{1.2}
\begin{tabular}{c|c|p{10cm}}
\Xhline{2.3\arrayrulewidth}
Feature Type & \multicolumn{1}{l|}{Direction} & Name \\ \hline
\multirow{3}{*}{timing-based} & forward & Flow \gls{IAT}\textsuperscript{*}, packets/sec \\ \cline{2-3} 
 & backward & Flow IAT\textsuperscript{*}, packets/sec \\ \cline{2-3} 
 & bi-direction & duration, Flow \gls{IAT}\textsuperscript{*}, packets/sec, bytes/sec, active time\textsuperscript{*}, idle time\textsuperscript{*} \\ \hline
\multirow{3}{*}{protocol-based} & forward & \# packets, packet length\textsuperscript{*}, PSH counts, URG counts, header length, initial TCP window size, avg segment size, subflow\textsuperscript{\dag} \\ \cline{2-3} 
 & backward & \# packets, packet length\textsuperscript{*}, PSH counts, URG counts, header length, initial TCP window size, avg segment size, subflow\textsuperscript{\dag},  \\ \cline{2-3} 
 & bi-direction & packet length\textsuperscript{*}, flag counts\textsuperscript{\S}, down/up ratio, protocol \\ \hline
ID-based & None & flow ID, src IP, dst IP, src port, dst port, timestamp \\ 
\Xhline{2\arrayrulewidth}
\end{tabular}
\egroup
\caption{Features used in \name. \textbf{*} means (min, max, avg, std) are computed for a given property. \textbf{\dag} indicates where (avg packets, avg bytes) are computed. \textbf{\S} indicates (FIN, SYN, RST, PSH, ACK, URG, CWE, ECE) are counted in flows.}
\label{table:features}
\end{table*}

\subsubsection{Feature Extraction}
\label{sec:feature_extraction}

\name employs a two-step process to extract numerical or categorical information (features) of the traffic observed, i.e., packet grouping and statistics computation. The former involves aggregating into flows packets generated between same pairs of applications, which can be achieved by monitoring origin, destination, and protocol fields. 

Since \name operates at the network layer and is not guaranteed to have access to packet payloads, we confine consideration to features that encompass timing statistics and protocol information. We find that employing popular open-sourced tools for feature extraction, such as CICFlowMeter \cite{flowmeter} (which should be able to extract 80+ statistical features), is problematic. 
Indeed, CICFlowerMeter uses a faulty mechanism to identify the end of TCP flows, which results in benign traffic often being mislabeled as malicious, and vice versa. 

\begin{figure}[]
    \centering
    \includegraphics[width=0.83\columnwidth]{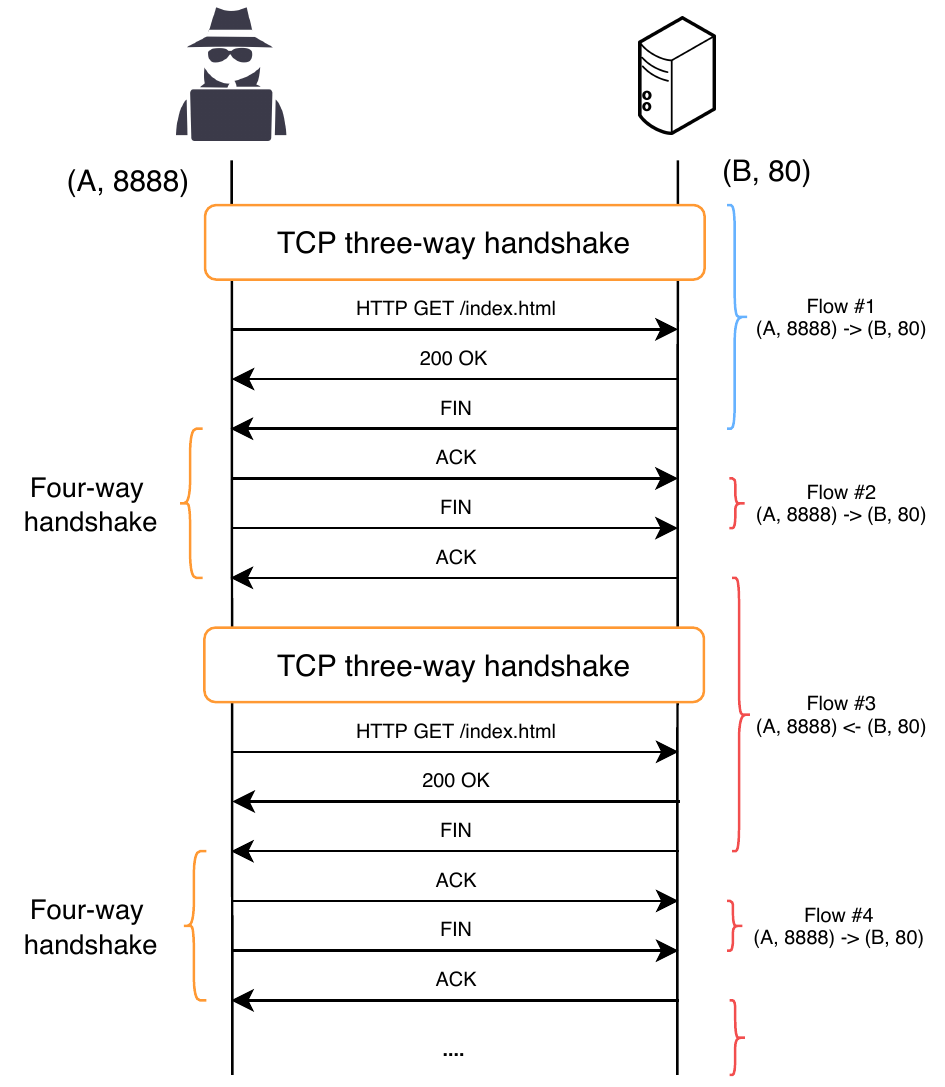}
    \caption{Incorrect flow labeling due to wrong TCP termination rules. Blue curly braces indicate the flow is labeled correctly; red curly braces indicate mislabeling.}
    \label{fig:flowgrouping}
\end{figure}

With the CICFlowMeter feture extractor, if a new incoming TCP packet has a FIN flag set, the packet is immediately deemed to be the last packet in that \textit{flow}. 
Obviously, this does not strictly follow the four-way handshake of TCP connection teardown. We show in Figure~\ref{fig:flowgrouping} that the premature assessment of termination leads to mislabeling, which is especially relevant to automated attacks such as \gls{DoS}. 

Assume that A is performing simple HTTP DoS attacks targeting B, quickly reusing a same port 8888. Also assume each time it is B who decides to terminate the TCP connections. Then, applying the mechanism described above on two consecutive flows from (A, 8888) to (B, 80) would generate 4 complete flows and an incomplete one,  
In this case, any flows from (A, 8888) to (B, 80) should be labeled as DoS. However, Flow \#2 and \#4 only consist of two packets (ACK and FIN) which cannot reflect any malicious purpose, while Flow \#3 that should be labeled as malicious, is marked benign because of its wrongly perceived direction. We confirm that this type of mislabeling occurs for DoS-Hulk attacks in the publicly available CSE-CIC-IDS-2018 dataset~\cite{ids2018} (which we use after correct relabeling), but further instances may exist.

We fix this logic error (along with other programming bugs encountered) in CICFlowMeter, by following a complete four-way handshake to terminate TCP flows. The timeout mechanism is preserved for stateless protocols. We also note that the original tool further extracts partial features that are not well-defined. Hence we revise the code and only output 69 features per flow, as summarized in Table \ref{table:features}.
\footnote{We will release our feature extractor's source code upon publication.}

\subsubsection{Sequence Generation}
Network anomaly is often nuanced, as a single network flow may be benign on its own, but observing multiple similar instances active at the same time may strongly suggest an automated attack in progress. Therefore, 
it is necessary to observe consecutive flows between hosts (applications) to further confirm malicious activity. 

As such, a \textbf{sequence} in \name is defined as successive flows using the same protocol between a pair of hosts, that is aggregated by (Src IP, Dst IP, Protocol). This is because with automated attacks,  many-to-one (DoS, brute forcing) and many-to-many (port scanning) port attacks between a pair of hosts are common. Therefore, we regard a sequence through grouping not by the tuple used for flow aggregation, but by the origin and destination addresses, along with  protocol type.

\setlength{\textfloatsep}{2pt}
\begin{algorithm}[t]
\small
  \caption{\name's sequence generation algorithm incorporating sliding windows and timeout thresholding.}
  \label{seq_gen}
  \begin{algorithmic}[1]
    \Inputs{Tiemout value $\tau$, and window size $\alpha$}
    \Initialize{\textit{conn\_table}: A connection table storing the incoming flows from the feature extractor. \\
    \textit{seq\_list}: A FIFO list buffering the inputs for the feature augmentor or the anomaly detector.}
    \While {True}
    \State {$start\_time$ $\gets$ $now()$}
    \For{$flow$ from the feature extractor}
    \State {$conn\_table[flow.id].add(flow)$}
    \If {$len(conn\_table[flow.id]) > \alpha$}
    \State {$seq\_list.add(conn\_table.remove(flow.id))$}
    \EndIf
    \If { $now() - start\_time > \tau$}
    \State {$seq\_list.addAll(conn\_table.removeAll())$ }
    \State {$start\_time$ $\gets$ $now()$}
    \EndIf
    \EndFor
    \EndWhile
    \end{algorithmic}
\end{algorithm}

We adopt a flexible approach to generating sequences, which is a combination of sliding window and timeout thresholding techniques, as described by Algorithm~\ref{seq_gen}. \name allows two user-defined parameters for this purpose, namely window size \(\alpha\) and timeout value \(\tau\), and maintains a connection table with two columns: \textit{ID} and \textit{seq\_list}. The \textit{ID} of each flow is a 3-tuple (Src IP, Dst IP, Protocol) and for each \textit{ID}, a FIFO \textit{seq\_list} is maintained, storing the flows with the same \textit{ID}. A newly generated flow is added to the \textit{seq\_list} determined by its \textit{ID}. Once the length of any \textit{seq\_list} is larger than \(\alpha\), the elements in the list are regarded as a sequence to be passed to the neural model. 
Meanwhile, after every \(\tau\) seconds, the entire connection table is emptied regardless of the length of the \textit{seq\_lists}. Any list whose length is less than \(\alpha\) is padded to \(\alpha\) for the purpose of alignment. 
This design is customizable: the larger \(\alpha\) is, the more comprehensive~the context that the ensemble model obtains, but the higher the memory requirements; the smaller \(\tau\) is, the more timely~clas\-sification can be achieved, at the cost of more compute resources. 

\subsubsection{Feature Augmentation} 
\label{sec:aug}
\begin{figure}[t]
\vspace*{-0.5em}
    \centering
    \includegraphics[width=0.92\columnwidth]{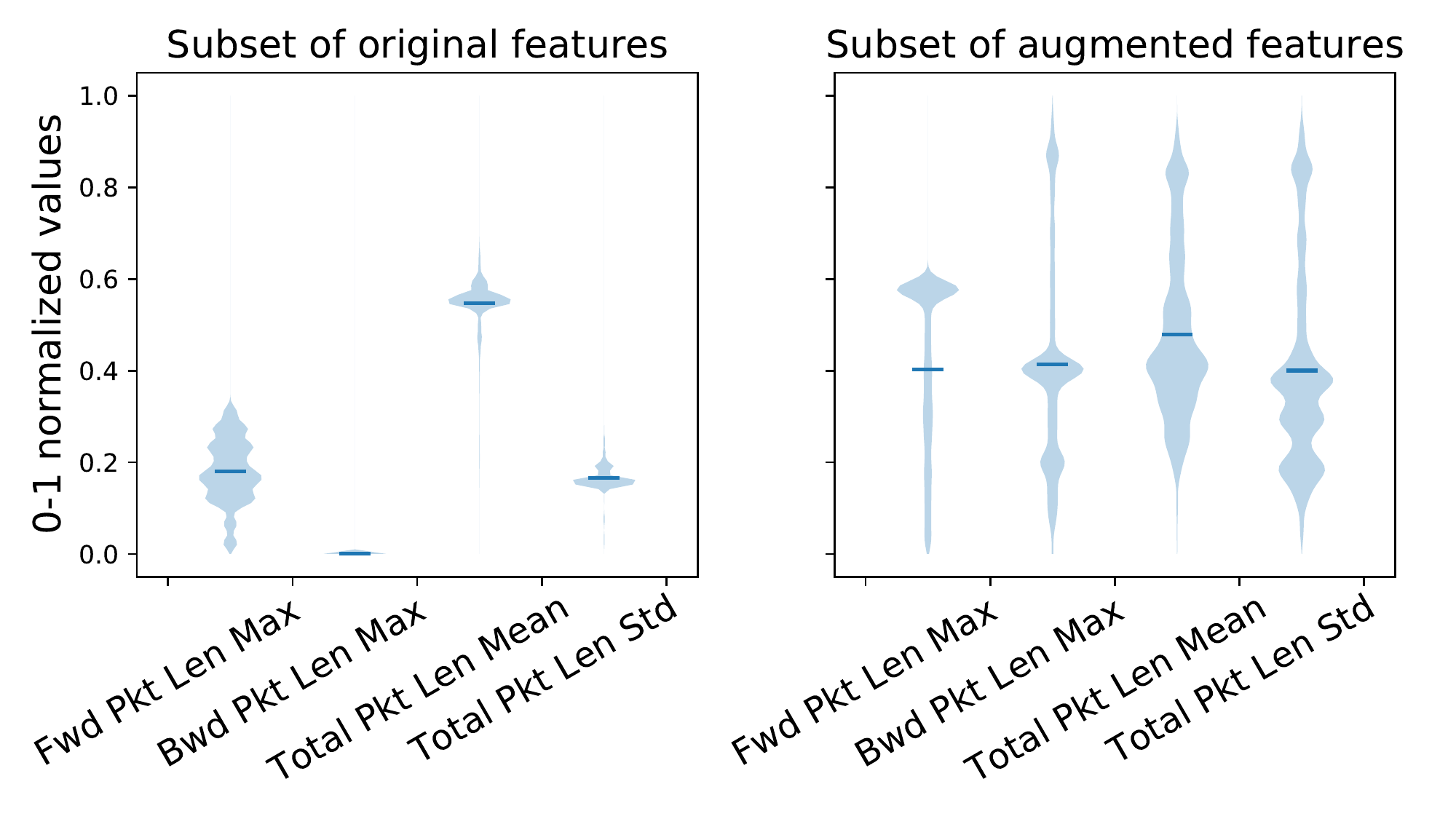}
    \vspace*{-1em}
    \caption{Violin plots of 4 features before and after data augmentation. All values are normalized between 0 and 1.}
    \label{fig:violin}
\end{figure}

Since data used for \gls{ML} training are largely collected in controlled environments, synthetically generated attacks may not offer an accurate view of network threats occurring in real-world~\cite{sommer2010}, which prevents the model from learning a reliable decision boundary. For example, a victim HTTP server was set up to produce the CSE-CIC-IDS2018 dataset \cite{ids2018}; during HTTP DoS generation, during HTTP DoS generation, all flows encapsulated the same backward payloads from victim to attacker, resulting in little variability in payload-related features (see Figure~\ref{fig:violin}, left). In reality, it is hard to predict how the victim would respond, and we show in \S\ref{sec:evaluation} that such artificially low variability leads to poor generalization abilities for a range of supervised models. 

We mitigate this problem by augmenting a collection of payload-related features to emulate a more realistic network environment. Specifically, we set up an HTTP victim server and an attacking client. The client only makes single requests with the \texttt{Keep-Alive} header over one TCP connection, to emulate HTTP \gls{DoS} attacks. The size of each request and response is sampled from two discrete uniform distributions:
 \setlength{\belowdisplayskip}{2pt} \setlength{\belowdisplayshortskip}{2pt}
 \setlength{\abovedisplayskip}{2pt} \setlength{\abovedisplayshortskip}{2pt}
\begin{align*}
    http\;request\;size &\sim \mathcal{U} (100, 400),\\
    http\;response\;size &\sim \mathcal{U} (100, 15000).
\end{align*}

\begin{figure}[t]
    \centering
    \includegraphics[width=\columnwidth]{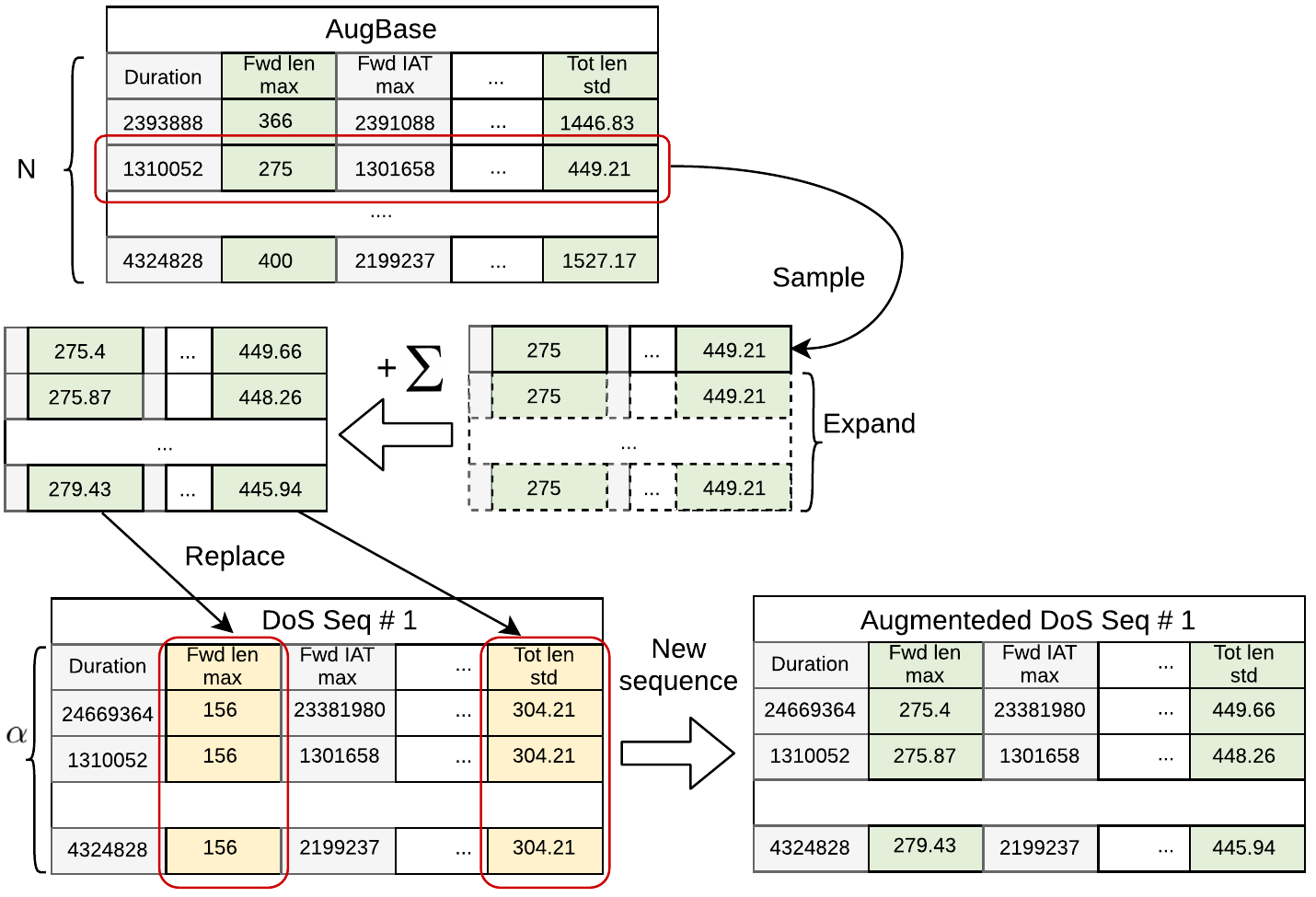}
    \caption{Illustration of the data augmentation process. Yellow cells denote payload features in original training data; green cells represents payload features from \texttt{AugBase}. $\Sigma$ is a noise matrix with each element sampled from $\sigma \sim \mathcal{N}(0, 5)$.}
    \label{fig:aug_data}
\end{figure}

In total, we generate 2,000 flows. 
A graphical illustration of the augmentation process is depicted in Figure \ref{fig:aug_data}. For each sequence that contains single-request HTTP DoS attacks (excluding Slowloris), (1) a flow is randomly sampled from the \texttt{AugBase} set and expanded to sequence length \(\alpha\); (2) random noise \( \sigma \sim \mathcal{N}(0, 5)\) is added to each payload-related cells to mimic minor differences among flows in a sequence; (3) finally, payload-related features in the original sequence are replaced by the new features generated at Step 2. By applying such augmentation, the new payload features of different flows in the same sequence would not differ much, but the features among different sequences would look considerably different. The distributions and the means of a subset of payload features in the augmented set are shown in Figure~\ref{fig:violin} (right). We use augmented data only for training. 

It is also worth noting that the augmented data need not originate from any real traffic, since we only change parts of the features. However, what the model can learn from the augmented set is that \emph{(i)} payload features possess high variability within some attacks (DoS in our case) whose logic does not rely on specific payloads, thus payload features should not be utilized for decision; and \emph{(ii)} the rest of features (payload-irrelevant) are more valuable in distinguishing augmented attacks. We \emph{choose not to remove payload-related features altogether because they may be important for the model to differentiate other types of attacks}, such as Slowloris, which only sends a small amount of payload during a long span. In \S\ref{sec:evaluation}, we demonstrate that augmentation boosts the performance of several supervised models.

\subsubsection{Ensemble Network}
The sequential ensemble neural network is the critical component of \name and is responsible for detecting malicious traffic based on the inputs provided by the sequence generator or feature augmentor. As explained in \S\ref{sec:defensive_scheme}, detecting automated cyber attacks is a time-sensitive task and hinges on temporal correlations between network flows. Let \(X:= \{\bm{x}_{1}, \bm{x}_{2}, ..., \bm{x}_{\alpha}\} \) and \( Y:= \{y_{1}, y_{2}, ..., y_{\alpha}\}\) denote a sequence of inputs and respectively their corresponding correct prediction. Then time-sensitive intrusion detection can be formalized as
\begin{align*}
\widetilde{\mathbf{\theta}} = \mathop{\arg\max}\limits_{\mathbf{\theta}}P_{\mathbf{\theta}}\left(y_{1}, y_{2}, ...,y_{a} | \bm{x}_{1}, \bm{x}_{2}, ..., \bm{x}_{a}\right),
\end{align*}
where the sequential model is parameterized by $\mathbf{\theta}$.

\name leverages two \gls{LSTM}-based models to approximate the probability function above. In the next section, we first introduce the LSTM models we employ, and show both conceptually and empirically that an ensemble of such sequential models, preferably with different architectures, is key to improving the overall \gls{NID} performance. 

\section{A Sequential Ensemble for NID}
\label{sec:ensemble}

We overview the different blocks that lay foundations for our Bi-ALSTMs, then explain the ensembling and why our approach is essential for high-performance classification.

\subsection{\glsfirst{LSTM}}
As a variant of \gls{RNN}, \gls{LSTM} \cite{hochreiter1997long} incorporates gating functions to simulate the update of memory units along time and has shown excellent ability to model long-term dependencies in sequential data \cite{sutskever2014sequence, kumar2022land}. An LSTM maintains two states: a cell state \( c_{t}\) and a hidden state \( h_{t}\), which is computed based on inputs up to timestamp \(t\), i.e., \( \mathbf{x}_{1}, ..., \mathbf{x}_{t} \;(\mathbf{x}_{i} \in \mathbb{R}^{d})\). To maintain long-term dependencies, the \gls{LSTM} cell has \textit{input}~(\( i_{t}\)), \textit{forget} (\( f_{t}\)), and \textit{output} (\( o_{t}\)) gates controlling the information flowing through at different timestamps. Gates are modeled by single-layer neural networks with parameters \(W_{x i}, W_{h i}, W_{x f}, W_{h f}, W_{x o}, W_{h o} \in \mathbb{R}^{h \times d} \) and  associated biases, i.e., 
\begin{flalign}
&i_{t} =\sigma\left(W_{x i} x_{t}+W_{h i} h_{t-1}+b_{i}\right),&\label{eq:lstm-i}\\
&f_{t} =\sigma\left(W_{x f} x_{t}+W_{h f} h_{t-1}+b_{f}\right),\\
&c_{t} =f_{t} \circ c_{t-1}+i_{t} \circ \tanh \left(W_{x c} x_{t}+W_{h c}, h_{t-1}+b_{c}\right), \label{eq:new_cell} \\
&o_{t} =\sigma\left(W_{x o} x_{t}+W_{h o} h_{t-1}+b_{o}\right),\\
&h_{t} =o_{t} \circ \tanh \left(c_{t}\right),\label{eq:lstm-o}
\end{flalign}
where \( \sigma\) denotes the sigmoid function and \( \circ\) represents element-wise product. When a new input \( x_{t}\) is given to the \gls{LSTM}, the cell state \( c_{t} \) is updated with information from \( x_{t} \) and the previous cell state \( c_{t-1}\). The proportion of \( x_{t}\) and \( c_{t-1}\) in the new cell state is determined by \( i_{t} \) and \( f_{t} \), as in Eq.~\ref{eq:new_cell}. \( h_{t}\) can be perceived as a non-linear transformations on \( c_{t} \), and are always used for downstream tasks, such as classification. In \name, an \gls{MLP} \(g_{\phi}(\cdot)\) is used to approximate the probability of \(y_{t}\) given \(h_{t}\).

When it comes to intrusion detection against automated network attacks, the order of attack flows in a sequence is less important.
Hence, at any timestamp \( t\), both previous inputs and subsequent inputs may be equally highly correlated with \( x_{t}\). \textit{The traditional \gls{LSTM} can only model temporal information unidirectionally as time evolves.} In other words, the hidden representation \(h_{t}\) only comprises the context before or at timestamp \( t\). To generate more comprehensive hidden representations for anomaly detection, a \gls{Bi-LSTM} which runs two \glspl{LSTM} separately (one forward,~one backward), and whose hidden states are concatenated before given to \(g_{\phi}(\cdot)\), can be used. The objective of \gls{Bi-LSTM} is thus:
\[
\begin{aligned}
&\max_{W, b, \bm{\phi}} p(y_{1}, y_{2}, ...,y_{a} | \bm{x}_{1}, \bm{x}_{2}, ..., \bm{x}_{a}, W, b, \bm{\phi}) \\
&= \max_{W, b, \bm{\phi}} \prod_{t=1}^{\alpha} p(y_{t}| \bm{x}_{1}, ..., \bm{x}_{\alpha}, W, b, \bm{\phi})
= \max_{W, b, \bm{\phi}} \prod_{t=1}^{\alpha} g_{\phi}(h_{t}),
\end{aligned}
\]
where \( W\) and \( b\) are the weights and biases in \gls{LSTM} cells and \(\bm{\phi}\) the parameters of the \gls{MLP}. \gls{Bi-LSTM} is one of the benchmarks we consider in evaluating our work.

\subsection{ConvLSTM}
\gls{ConvLSTM} \cite{xingjian2015convolutional} was first proposed to model spatiotemporal data, such as radar echo maps, whose spatial correlations cannot be extracted by fully connected layers in \gls{LSTM}. Conv\-LSTM tailors the convolution operation into \gls{LSTM} by replacing matrix multiplication operations in (\ref{eq:lstm-i})--(\ref{eq:lstm-o}) with convolution, as follows:
\begin{flalign}
&i_{t} =\sigma\left(W_{x i} * \mathcal{X}_{t}+W_{h i} * \mathcal{H}_{t-1}+b_{i}\right),&\\
&f_{t} =\sigma\left(W_{x f} * \mathcal{X}_{t}+W_{h f} * \mathcal{H}_{t-1}+b_{f}\right),\\
&\mathcal{C}_{t} =f_{t} \circ \mathcal{C}_{t-1}+i_{t} \circ \tanh \left(W_{x c} * \mathcal{X}_{t}+W_{h c} * \mathcal{H}_{t-1}+b_{c}\right) \hspace*{-1em}\\
&o_{t} =\sigma\left(W_{x o} * \mathcal{X}_{t}+W_{h o} * \mathcal{H}_{t-1}+b_{o}\right),\\
&\mathcal{H}_{t} =o_{t} \circ \tanh \left(\mathcal{C}_{t}\right),
\end{flalign}
in which \( *\) denotes the convolution operator, $ \mathcal{X}_{t}\in \mathbb{R}^{d \times 1}$ is the input, and $W_{x i}, W_{h i}, W_{x f}, W_{h f}, W_{x o}, W_{h o} \in \mathbb{R}^{k \times c}$ are the convolution kernels. The above applies to a single Conv\-LSTM unit, which can be further extend to multiple layers as traditional \gls{LSTM} does. 
A pooling layer can be added between each two \gls{ConvLSTM} layers to reduce computation.

An immediate concern is whether applying convolution-embedded models on network traffic data without obvious spatial information would be effective. In fact, \gls{CNN} not only gained success in computer vision \cite{he2016deep, yang2021msta}, but also in areas including web traffic fingerprinting \cite{deepfp} and mobile traffic forecasting~\cite{zhang:2019}. As the sequential data we deal with are one-dimensional, we implement a 1-D \gls{ConvLSTM}, which takes inputs with channel and length dimensions, where channel by default equals 1. 


\textbf{Differences from CNN-LSTM.} The CNN-LSTM, or \gls{LRCN}, is a combination of CNN and LSTM, first proposed for visual recognition. It differs from \gls{ConvLSTM} in that in the former a separate \gls{CNN} handles spatial information before providing input to \gls{LSTM}. In contrast, the latter has a compact form. While previous studies apply CNN-LSTM to \gls{NID} \cite{jiang2020network}, our work is the first to leverage \gls{ConvLSTM} and study the differences between the two structures. Our results in \S \ref{sec:evaluation} reveal that \gls{ConvLSTM} consistently outperforms CNN-LSTM. 

\subsection{Bidirectional Asymmetric LSTM}
\gls{Bi-LSTM} and Bi-\gls{ConvLSTM} can be perceived as ensemble models with two separate \gls{LSTM} units. The hidden states from two units are concatenated so that future information is accessible at the current timestamp \( t\), which can potentially benefit the downstream classification task. Since our targets are sequences with \textit{similar} malicious flows, it is the hidden states at the two ends of the structures that can acquire most information. The hidden representations in the middle of a sequence from two (Conv-)\gls{LSTM} units yield certain amount of redundant information when the same architecture is used.

\setlength{\textfloatsep}{5pt}
\begin{algorithm}[!b]
\small
  \caption{The training algorithm for \gls{Bi-ALSTM}}
  \label{bialstm_training}
  \setstretch{1.2}
  \begin{algorithmic}[1]
    \Inputs{\mbox{$\mathcal{D} := \{S_{1},...,S_{n}\}; S_{i} := \{ (\mathbf{x}_{i1}, y_{i1}),...,(\mathbf{x}_{i\alpha}, y_{i\alpha})\},$} \\
    $\lambda_{1} := L_{2}\; weight\;decay, \; \lambda_{2} := learning\;rate.$ }
    \Initialize{ Denote $h_{W_{fc}}(\cdot)$ and $h_{W_{conv}}(\cdot)$ as the \gls{LSTM} and \gls{ConvLSTM} unit parameterized by $W_{fc}$ and $W_{conv}$, and predictor $g_{\phi}(\cdot)$ parameterized by $\phi$. $W_{fc}$, $W_{conv}$, $\phi$ set via Xavier initialization~\cite{glorot2010understanding}. }
    \While {model has not converged}
    \For{$S_{i}$ sampled from $\mathcal{D}$}
    \State {$\overrightarrow{X_{i}} \gets (\mathbf{x}_{i1}, ..., \mathbf{x}_{i\alpha} ), \overrightarrow{\mathbf{y}_{i}} \gets (y_{i1}, ..., y_{i\alpha}) $}
    \State {$\overleftarrow{X_{i}} \gets reverse(\overrightarrow{X_{i}})$}
    \State {$\overrightarrow{H}_{i, fc} \gets h_{W_{fc}}(\overrightarrow{X}_{i})$}
    \State {$ \overrightarrow{H}_{i, conv} \gets reverse(h_{W_{conv}}(\overleftarrow{X}_{i})) $}
    \State {$ \overrightarrow{H}_{i} \gets \overrightarrow{H}_{i, fc} \oplus \overrightarrow{H}_{i, conv} $} \Comment{\parbox[t]{.3\linewidth}{Eq. \ref{eqn:combine}}}
    \State {$\mathcal{L} \gets NLL(softmax(g_{\phi}(\overrightarrow{H}_{i}), \overrightarrow{\mathbf{y}_{i}}) + $ 
    \Statex \qquad \qquad $\lambda_{1}(||W_{conv}||_{2}^{2} + ||W_{fc}||_{2}^{2} + ||\phi||_{2}^{2})$}
    \State {$
        \phi, W_{conv}, W_{fc} \gets Adam(\mathcal{L}, \phi, W_{conv}, W_{fc}; \lambda_{2})$}
    \EndFor
    \EndWhile
    \State {\Return {$W_{fc}, W_{conv}, \phi$}}
    \end{algorithmic}
\end{algorithm}

Given the fact that different architectures are likely to exploit different facets of input features for classification \cite{raghu2021vision}, we propose \textbf{\glsfirst{Bi-ALSTM}}, which consists of two different, asymmetric \gls{LSTM} units, one for forward and the other for backward processing, to generate intermediate representations that incorporate more comprehensive temporal contexts. The hidden states from two different units are first linearly combined, then fed through an activation function. Precisely, denote \(\mathbf{h}_{fc}^{t} \in \mathbb{R}^{N_{1}} \) as the hidden state generated by \gls{LSTM} at timestamp \(t\), and \( \mathbf{h}_{conv}^{t} \in \mathbb{R}^{N_{2}}  \) generated by \gls{ConvLSTM} (operated backwards). The final hidden states are formed through the following fusion operation:
\begin{equation}
\label{eqn:combine}
\mathbf{h}^{t} = \tanh\left( \frac{\mathbf{U}_{conv}\mathbf{h}_{conv}^{t}}{||\mathbf{U}_{conv}\mathbf{h}_{conv}^{t}||_{2} } +  \frac{\mathbf{U}_{fc}\mathbf{h}_{fc}^{t}}{||\mathbf{U}_{fc}\mathbf{h}_{fc}^{t}||_{2} }\right), 
\end{equation}
where \( \mathbf{U}_{conv} \in \mathbb{R}^{N \times N_{2}}\) and \( \mathbf{U}_{fc} \in \mathbb{R}^{N \times N_{1}}\) are learnable parameters. \(\mathbf{h}_{fc}^{t}\) and \(\mathbf{h}_{conv}^{t}\) are projected to the same subspace and \(L_{2}\)-normalized, so that in the learning process, any single unit would not easily dominate the final results. This design allows two asymmetric \glspl{LSTM} to produce hidden states with different dimensions, meaning that two
different \gls{LSTM} structures can be tuned separately and flexibly before building the \gls{Bi-ALSTM}. 
Finally, a single FC layer \(g_{\phi}(\cdot): \mathbb{R}^{N} \rightarrow \mathbb{R}^{C}\) with softmax function is used to approximate the probability of the samples belonging to each class. 

\textbf{Computational complexity}:  We first derive the time complexity of a single-step forward pass of LSTM. Given that the time complexity of $W_{x}x_{t}$ is $O(hd)$, it is easy to know that the time complexity inside each nonlinearity $\sigma$ is $O(hd+h^{2}) = O(h(d + h))$. Assuming $\sigma$ and $tanh$ have a constant time complexity, applying $\sigma$ or $tanh$ element-wise yields the complexity of $h$, which can be omitted due to the existence of $O(h^{2})$. Therefore, the time complexity of a single-step forward pass equals $O(h(d+h))$. Similarly, given that $W_{h} * \mathcal{X}_{t}$ has time complexity $O(dkc$), the time complexity of a single-step forward pass of ConvLSTM is $O(dkc + dkc^{2}) = O(dkc^{2})$. Assume the input sequence to Bi-ALSTM has the length $n$. The computational complexity of a single-layer Bi-ALSTM results in $O(n(h(d+h)+dkc^{2}))$.

We use negative log likelihood with \(L_{2}\) regularization as the loss function. \gls{Bi-ALSTM} is stochastically optimized via back propagation through time by the Adam algorithm. The complete training process follows Algorithm \ref{bialstm_training}.

\subsection{Why an All-range Multi-class Classifier is Unfeasible}
Applying supervised methods for \gls{NID} commonly involves training a multi-class classifier that seeks to detect as many types of malicious attacks as possible. However, we argue that this approach would lead to an overfitted model because the \textit{ambiguity of the true attack labels} is widely overlooked. To understand this, consider a classifier is trained to differentiate two types of network attacks: SQLmap fuzzing vs. Web password Bruteforcing,
%
both of which repetitively initiate HTTP requests to a given API. Although the two types of attacks would incorporate different payloads (SQL scripts and user/password pairs respectively), this discrepancy appears negligible on a statistical level, because the contents of  payloads would not be extracted. Given that both trigger database I/O operation, the extracted timing information would be hardly differentiable. 

Thus if a trained model can distinguish between such attacks, it has to be overfitted and learn a decision boundary that~is unique to the dataset, rather than truly understand the differences. Attack labels usually indicate the purposes and techniques behind them, but when it comes to network layer, the attack realizations, i.e., traffic flows, would not be clearly dissimilar. In this regards, using sequential models to distinguish as many types of attacks as possible is unrealistic. 

\subsection{Abstract Labeling}
Given that it is hard to correctly predict every type of automated cyber attack with a network-based algorithm, reverting to a binary detector seems sensible. However, since we augment DoS attacks, we decide to use \textbf{abstract labeling} in order to evaluate the augmentation technique and to avoid the aforementioned overfitting issue.
%
Specifically, we assign a number of abstract, generic labels, including \textit{benign}, \textit{DoS}, \textit{portscanning} and \textit{bruteforcing \& fuzzing}, as ground truth during training. 
On one hand, this approach can clearly illustrate the influence of our feature augmentation technique.
On the other hand, the model would not put effort in distinguishing the subtle differences between, e.g., DoS HOIC and DoS LOIC, which may not be separable by a network-based algorithm. 


\section{Experiments}
\label{sec:evaluation}

We implement \name in PyTorch and train the model on a GeForce Titan X GPU. To build the \gls{Bi-ALSTM}, we use an \gls{LSTM} unit for the forward pass and a {ConvLSTM} unit for the backward pass. The \gls{LSTM} unit has the following structure: $dropout(0.5) \rightarrow lstm(65, 48) \rightarrow lstm(48, 48)$, where the arguments in $lstm(\cdot, \cdot)$ denote the input size and the hidden size. The \gls{ConvLSTM} unit encompasses $convlstm1D(1, 3, 3) \rightarrow convlstm1D(3, 6, 3) \rightarrow maxpool()$, in which the arguments in $convlstm1D(\cdot, \cdot, \cdot)$ represent the input channel size, output channel size, and kernel size. The fused hidden states are passed through: $dropout(0.3) \rightarrow MLP(32, 5)$. The $L_{2}$ penalty and learning rate are set to 0.5 and 0.001 respectively.

For the Flow Aggregator/Sequence Generator, apart from our TCP termination fix (see \S\ref{sec:feature_extraction}), we set flow timeout to 30s and subflow duration 5s in CICFLowMeter. The Sequence Generator uses  timeout $\tau = 30$s and window size $\alpha=10$.

\subsection{Datasets}
We experiment with two datasets published by the \gls{CIC}, as described below. 

\textbf{CIC-IDS-2017} \cite{sharafaldin2018toward} contains most common cyber attacks, including bruteforcing, heartbleed, botnet, (D)DoS, Infiltration, and Web attacks. Traces were collected in a LAN, with benign traffic generated by profiling normal online behaviors of 25 users on different \glspl{OS}, including Win Vista, Win 7, Win 8, Win 10, Mac OS, and Ubuntu~12.
Attacks are produced by 4 different machines with one running Kali Linux and three Win 8.1. The traffic
collection spans 5 working days, and in total 51.1 GB of pcap files are open-sourced.
The feature sets of the corresponding {\tt pcap} files were published, but as we reveal in Section \ref{sec:feature_extraction}, these were extracted incorrectly. Hence, we only use the raw capture files in our experiments.

\textbf{CSE-CIC-IDS2018} \cite{ids2018} was generated 
in a much larger environment where an organizational LAN with five subnets simulated give different departments.  
450 machines act as normal users and 50 as attackers. The dataset contains a wider range of benign traffic, including HTTPS, HTTP, SMTP, POP3, IMAP, SSH, and FTP, and contains a larger attack collection (17 types). The dataset spans 10 days. 

\textbf{Self-collected traffic} -- since \textit{FTP-Bruteforce} and \textit{DoS-SlowHTTP\-Test} attacks were erroneously collected in CSE-CIC-IDS2018, 
the generated traffic merely contains SYN and RST packets. To mitigate this, we collected FTP-brute\-forcing traffic ourselves, generating 4,050 flows. We decide not to collect \textit{DoS-SlowHTTPTest} traffic, since a similar type of attack, i.e., \textit{DoS-Slowloris}, exists in CSE-CIC-IDS2018. 
We remove the mislabeled traffic and merge the self-collected traffic with the CSE-CIC-IDS2018 dataset. In the rest of our paper, we use CSE-CIC-IDS2018 to refer to the merged dataset. We employ our revised version of CICFlowMeter to generate network flows based on the {\tt pcap} files. The statistics of both datasets are shown in Table \ref{table:dataset}. 

\begin{table}[h]
\small
\centering
\def\arraystretch{1.2}
\begin{tabular}{p{1.35cm}p{1.3cm}p{1.2cm}p{1.2cm}p{1.5cm}}
\Xhline{2\arrayrulewidth}
 & \# \mbox{Features} & \# Instances & Anomaly ratio & Automated attack ratio \\ \hline
IDS-2017 & 69 & 2,607,289 & 0.2 & 0.189 \\
IDS-2018 & 69 & 8,786,169 & 0.1806 & 0.1802 \\
\Xhline{2\arrayrulewidth}
\end{tabular}
\caption{Statistics of datasets used for experimentation. ID-based features in Table \ref{table:features} not included in the training; `protocol' is hot-encoded. Thus, only 65 features used in total.}
\label{table:dataset}
\end{table}


\subsection{Cross-Evaluation}
We adopt a rigorous evaluation methodology, aiming to show the true generalization ability of our design, by \textbf{cross-evaluation}. Normally, a dataset is split into training and testing subsets, and the results on the test set compared across different algorithms. However, network environments are heterogeneous and data collected in one environment may not accurately reflect the diversity seen in practice. To test if an algorithm can truly distinguish the same type of malicious traffic in a different network topology, we also evaluate it on a second, unseen dataset. CSE-CIC-IDS2018 is split into training (70\%) and test (30\%) sets. To maintain time consistency, training data is selected from events that took place before the test data.
CIC-IDS-2017 is purely used for cross-evaluation, after the model was trained on the former. 

\begin{table*}[t]
\footnotesize
\centering
\bgroup
\def\arraystretch{1.2}
\begin{tabular}{c|ccc|ccc}
\Xhline{2\arrayrulewidth}
\multirow{2}{*}{Algorithm} & \multicolumn{3}{c|}{CSE-CIC-IDS2018} & \multicolumn{3}{c}{CIC-IDS-2017 (X-eval)} \\
 & precision & recall & F1 & precision & recall & F1 \\ \hline
RIPPER & 0.9983 & 0.0981 & 0.1786 & 0.0873 & 0.0106 & 0.0190 \\
Decision Tree & 0.9989 & 0.9990 & 0.9990 & 0.5385 & 0.3717 & 0.4398 \\
MLP & 0.9989 & 0.9962 & 0.9976 & 0.6736 & 0.4631 & 0.5435 \\
CNN & 0.9947 & 0.9951 & 0.9949 & 0.7705 & 0.6344 & 0.6958 \\
Autoencoder & 0.7783 & 0.7500 & 0.7639 & 0.4362 & 0.4197 & 0.4278 \\
OC-NN & 0.9722 & 0.5310 & 0.6868 & 0.7844 & 0.5136 & 0.6208 \\
Kitsune & 0.6310 & 0.6081 & 0.6193 & 0.4086 & 0.3932 & 0.4007 \\
DAGMM & 0.8666 & 0.8253 & 0.8454 & 0.4159 & 0.3116 & 0.3576 \\
Bi-LSTM & 0.9990 & 0.9979 & 0.9985 & 0.7258 & 0.4209 & 0.5317 \\
CNN-Bi-LSTM& \textbf{0.9996} & 0.9982 & 0.9989 & 0.8813 & 0.3750 & 0.5261 \\ 
Bi-ConvLSTM & 0.9984 & 0.9971 & 0.9977 & 0.8721 & \textbf{0.9693} & 0.9178 \\
Bi-ALSTM & 0.9994 & \textbf{0.9990} & \textbf{0.9992} & \textbf{0.9116} & 0.9446 & \textbf{0.9275}  \\ \hline
\multicolumn{1}{p{1.75cm}|}{Algorithm + } & \multicolumn{3}{c|}{CSE-CIC-IDS2018} & \multicolumn{3}{c}{CIC-IDS-2017 (X-eval)} \\
\multicolumn{1}{p{1.75cm}|}{augmentation} &
\multicolumn{1}{p{0.8cm}}{precision} & recall & \multicolumn{1}{p{0.65cm}|}{F1} & precision & recall & F1 \\ \hline
RIPPER & 0.9980 & 0.0934 & 0.1709 & 0.4998 & 0.1837 & 0.3687 \\
Decision Tree & 0.9989 & \textbf{0.9993} & \textbf{0.9991} & 0.5897 & 0.8556 & 0.6914 \\
MLP & 0.9989 & 0.9963 & 0.9976 & 0.7540 & 0.8690 & 0.8071 \\
CNN & 0.9925 & 0.9847 & 0.9886 & 0.7453 & 0.8687 & 0.8021 \\
Bi-LSTM & 0.9991 & 0.9956 & 0.9973 & 0.8555 & 0.9777 & 0.9125 \\
CNN-Bi-LSTM & 0.9996 & 0.9966 & 0.9981 & 0.8479 & 0.9683 & 0.9041 \\ 
Bi-ConvLSTM & 0.9996 & 0.9975 & 0.9985 & 0.8728 & 0.9780 & 0.9222 \\
Bi-ALSTM & \textbf{0.9997} & 0.9976 & 0.9987 & \textbf{0.9190} & \textbf{0.9800} & \textbf{0.9485} \\
\Xhline{2\arrayrulewidth}
\end{tabular}
\egroup
\caption{Precision, recall, and F1 score for Bi-ALSTM and benchmarks. NB: only supervised algorithms can be trained with augmented data and only HTTP (D)DoS attacks are augmented; semi-supervised methods only require benign traffic for training.}
\label{table:overall_results}
\end{table*}

\subsection{Benchmarks}

For comparison, we implement a range of benchmarks, including basic ML/DL structures (\gls{MLP}, \gls{CNN}, autoencoder, RIPPER \cite{lee1998data}, Decision Tree); state-of-the-art anomaly/ intrusion detectors, i.e., \gls{OC-NN}~\cite{ruff2018deep}, Kitsune/KitNET \cite{mirsky2018kitsune} and \gls{DAGMM} \cite{zong2018deep}; and three \gls{Bi-LSTM} \mbox{variants~\cite{jiang2020network}.} 

\gls{OC-NN} \cite{ruff2018deep} and \gls{DAGMM} \cite{zong2018deep} are offline semi-supervised algorithms for general anomaly detection. 
\gls{OC-NN} aims to learn a mapping for the benign samples to a kernel space where the majority of them can be enclosed by a hypersphere. 
During the testing phase, the distances from the samples to the center of the hypersphere represent the anomaly score of the data. 
Different from \gls{OC-NN}, \gls{DAGMM} models the benign data from a probabilistic perspective with a mixture of Gaussian distributions. 
The negative probability of the data being sampled from the \gls{PDF} represents the anomaly score. 

Kitsune \cite{mirsky2018kitsune} is an online semi-supervised \gls{NIDS}. It uses an ensemble of shallow autoencoders to learn the features of benign data in different subspaces; a final autoencoder fits the correlations of the reconstruction errors from the shallow autoencoders. 
The neural architecture is named KitNET. 
During testing, the reconstruction errors are computed to represent the degree of abnormality. For a fair comparison, KitNET is trained in an offline manner with more than one epoch. 

For semi-supervised algorithms (\gls{OC-NN}, \gls{DAGMM}, KitNET and Autoencoder), an anomaly ratio \( \alpha \) needs to be preset, indicating the proportion of anomalous samples, and during the testing phase, the data with the top \( \alpha \times 100\% \)
of the anomaly scores are classified as anomalous. The anomaly ratio is set to 0.189 and 0.1802 on CIC-IDS-2017 and CSE-CIC-IDS2018 respectively, which is the same percentage of automated attacks in the two datasets.

RIPPER \cite{lee1998data} and Decision Tree are two basic machine learning models, where the first one aims to generate a simple ruleset for classifications while the second embeds rules in a tree by recursively finding the best splits. 

The structures of \gls{Bi-LSTM} and Bi-\gls{ConvLSTM} resemble the units in \gls{Bi-ALSTM}. For CNN-Bi-LSTM, an extra \gls{CNN} block, with the structure:
$Conv1D(1, 3, 3)$ $\rightarrow$ $MaxPooling$ $\rightarrow$ $Conv1D(3, 6, 3)$ $\rightarrow$ $MaxPooling$, is implemented. The arguments in $Conv1D(\cdot, \cdot, \cdot)$ represent input channels, output channels, and kernel sizes.

Prior to testing, we retrain all benchmarks with all the features in the CSE-CIC-IDS2018 dataset, which is richer than the datasets used for training in the original papers.

\subsection{Evaluation Metrics}
The average precision, recall and F1 score are commonly used to evaluate the performance of anomaly detection algorithms. These metrics can be measured based on the \gls{TP}, \gls{FP}, \gls{TN} and recall are computed as
$precision$ $=$ $TP/(TP+FP)$, $recall$ $=$ $TP/(TP+FN)$.
The precision indicates how likely the algorithm would give true alarms, and the recall measures how sensitive the algorithm is towards anomalies. There exists a trade-off between precision and recall, and to obtain an overall performance measure, their harmonic average is computed, i.e., the F1 score:
$F1 = 2 \times \frac{precision \;\times\; recall}{precision \;+\; recall}$.

We do not measure accuracy i.e., the percentage of the correctly classified samples, which is unlikely to reveal the algorithms' true NID performance: consider a dataset with 80\% benign and 20\% malicious instances; a model that classifies everything as benign has the same accuracy as a model that correctly recognizes all but 20\% of the benign~traffic.

For the ML-based algorithms that output an anomaly score for each test instance rather than just the predicted class, system administrators may choose a threshold higher than 0.5, which guarantees that the classifier has a lower \gls{FPR}. We plot the \gls{ROC} curve for sequential models, to evaluate their performance when the anomaly threshold is varied in $[0, 1]$. The \gls{ROC} curve is obtained by plotting the \gls{TPR} against \gls{FPR}. The closer to 1 the \gls{AUC} is, the better the classifier performs.

Given that our datasets consist of multiple types of cyber attacks, we further plot the \gls{ECDF} of the anomaly score with respect to each type of traffic on the crossed evaluated dataset, to illustrate the confidence of each sequential model.

Beyond the metrics for classification performance, we also care about the computational overhead of our design, and therefore report \gls{MACs}, the number of parameters and the concurrent processing capacity of each model on a edge GPU. \gls{MACs} and the parameter numbers reveal the complexity of different algorithms at a micro level, while the concurrent processing capacity on GPU can reflect the computational bottlenecks.

\subsection{Performance without Augmented Data}
We summarize our comparison in terms of threat detection performance between our Bi-ConvLSTM/-ALSTM models and the benchmarks considered, in Table~\ref{table:overall_results}. In the upper half, the different algorithms are trained on non-augmented data. The performance of the benchmark algorithms and those adopted by our \name is similar on the CSE-CIC-IDS2018 dataset, most of them attaining average metrics above 0.99. CNN-Bi-LSTM slightly outperforms other algorithms in terms of precision, while \gls{Bi-ALSTM} yields the highest recall and F1 score. An interesting finding is that the semi-supervised algorithms for general anomaly detection may not be suitable for network intrusion detection. Autoencoder, OC-NN and DAGMM cannot compete with basic supervised ML algorithms. One of the core assumptions for semi-supervised anomaly detection is that the algorithm can learn the characteristics of benign data, by estimating the probability, reconstructing the benign samples or finding an appropriate hyper-boundary enclosing them. However, network traffic is highly heterogeneous, serving with different protocols various applications, such as email, web browsing, streaming, etc. 
It remains questionable whether the aforementioned assumption holds on such a large range of `benign data'. Besides, detecting malicious traffic, especially automated attacks (which is our objective), appears to be a time-sensitive task. Therefore, observing a single instance may be insufficient to make reliable decisions. Consequently, existing anomaly detection algorithms tend to ignore this, which leads to modest results. 

\begin{figure}[b]
    \centering
    \includegraphics[width=0.95\columnwidth]{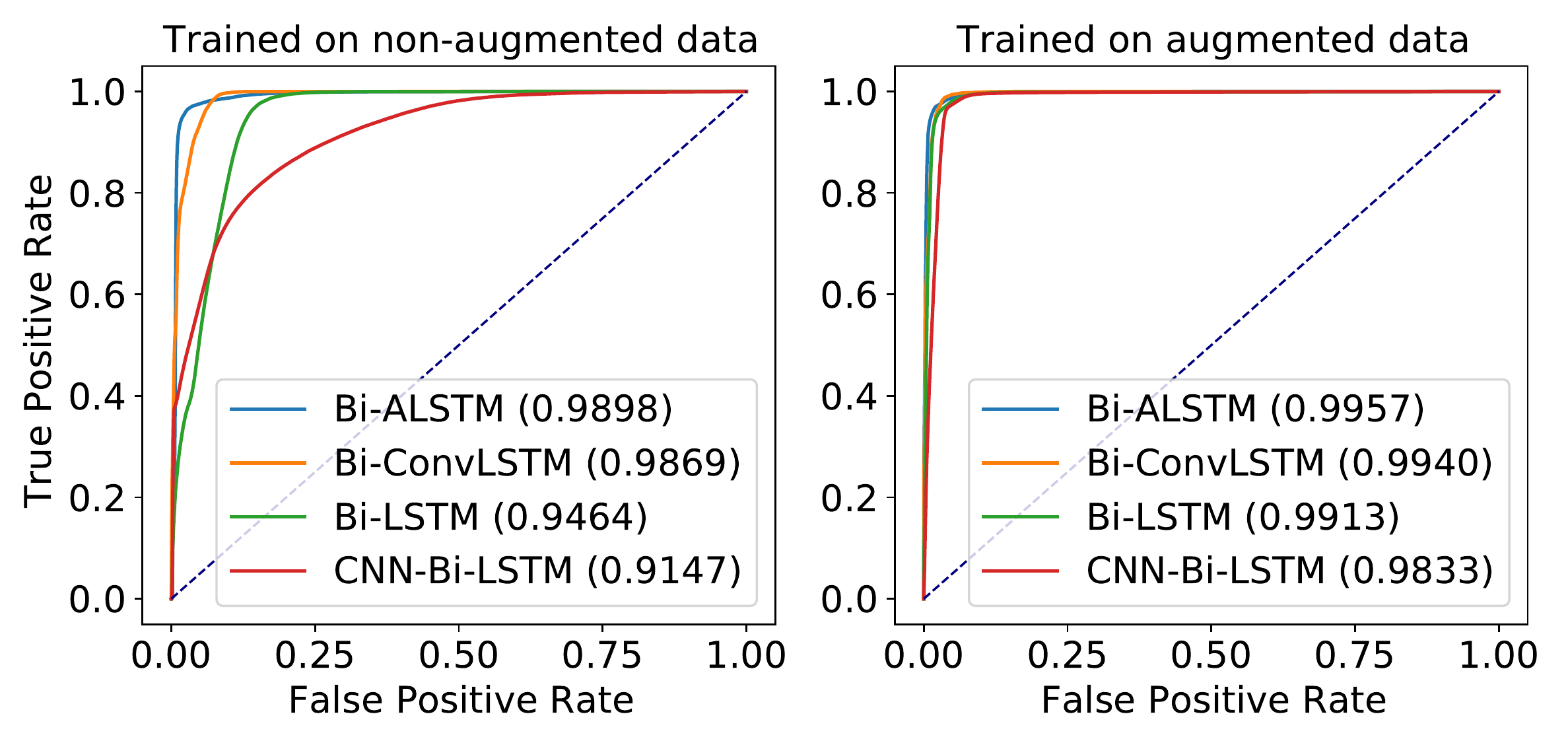}
    \caption{\gls{ROC} curves of \gls{LSTM}-based algorithms. \gls{AUC}  behind~labels. Models trained w/o (left) and w/ augmented data (right).}
    \label{fig:roc_auc}
\end{figure}

\begin{figure}[t!]
    \centering
    \includegraphics[trim=10 10 10 10 10 clip,height=0.9\columnwidth,angle=270]{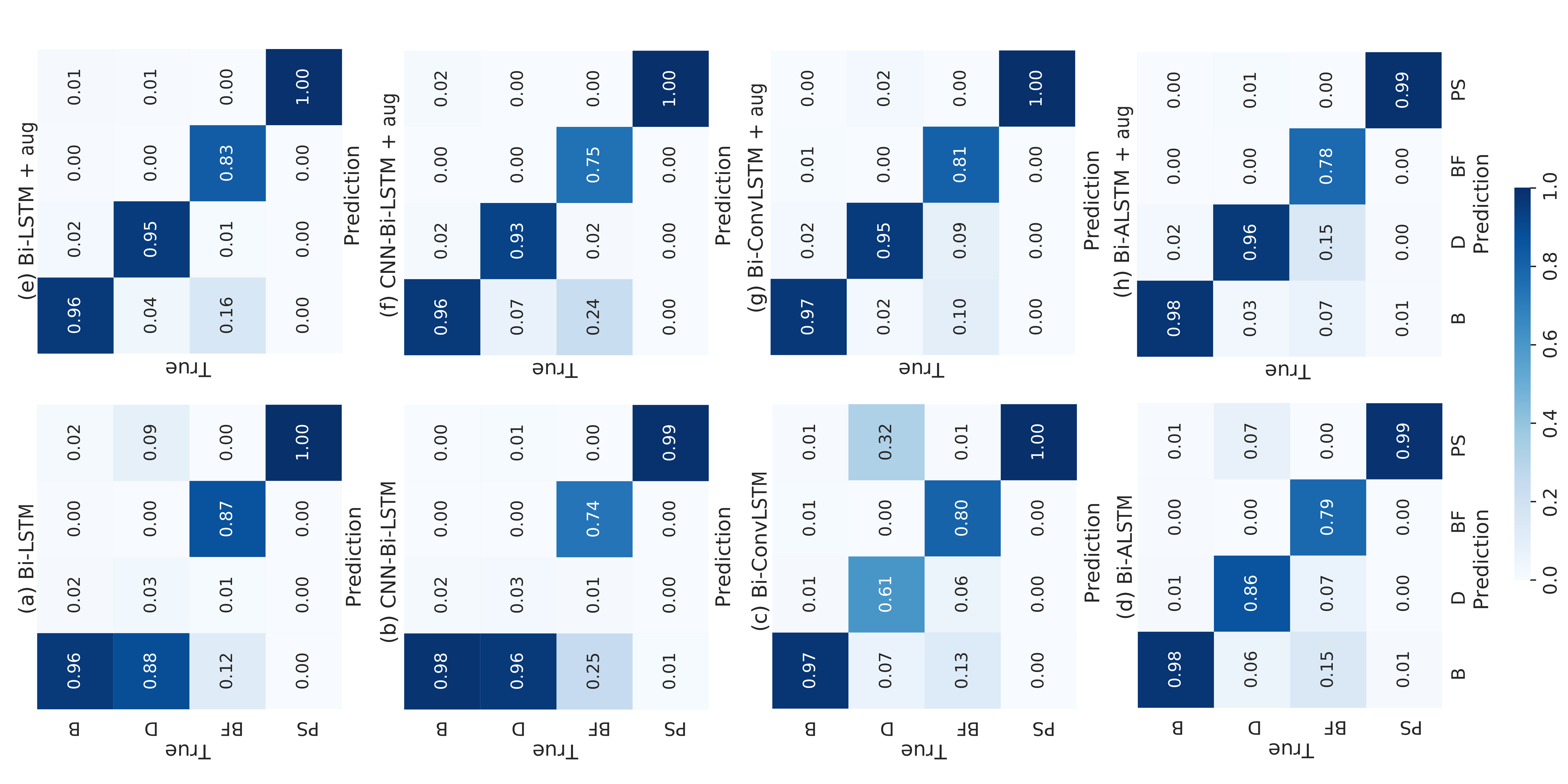}
\caption{Normalized confusion matrices (row values add to 1) for LSTM-based models cross-evaluated on CIC-IDS-2017. Models trained with/without augmented dataset (left/right). B represents \textbf{B}enign, D \textbf{D}oS, BF \textbf{B}ruteforcing and \textbf{F}uzzing, and PS \textbf{P}ort\textbf{S}caning. Numbers on diagonals are recalls. 
}
\label{fig:confuxion}
\end{figure}

\emph{The advantage of Bi-ConvLSTM and \gls{Bi-ALSTM} can be clearly seen on cross-evaluation results, where our models maintain consistently competitive performance.} Both attain F1 scores above 90\%, while other supervised algorithms, including \gls{Bi-LSTM} and CNN-Bi-LSTM exhibit a significant performance drop (F1 score around 50\%). We notice that though both CNN-LSTM and \gls{ConvLSTM} are proposed to handle spatiotemporal data, there is an obvious difference in performances, both in terms of F1 score (Table~\ref{table:overall_results}) and \gls{ROC} (Figure~\ref{fig:roc_auc}) on intrusion detection. As shown in Figs~\ref{fig:confuxion} (a), (b), \gls{Bi-LSTM} and CNN-Bi-LSTM do not learn a reliable decision boundary between benign traffic and \gls{DoS} attacks without the augmented data. Bi-ConvLSTM clearly outperforms them, yet still exhibits a high probability of classifying \gls{DoS} as port scanning attacks, as illustrated in Figure~\ref{fig:confuxion}(c), whereas \gls{Bi-ALSTM} is the most reliable (Figure~\ref{fig:confuxion}(d)).  

\subsection{Performance with Augmented Data}
The data augmentation procedure we propose is highly effective in helping the models generalize well. \emph{When the supervised models are trained with the augmented dataset, a remarkable performance gain can be observed in the cross-evaluation results} (Table~\ref{table:overall_results}, bottom half). The F1 scores of the benchmarks increase by at least 16\%, and \gls{Bi-LSTM} and CNN-Bi-LSTM even jump to 90\%. The improvements of Bi-ConvLSTM and \gls{Bi-ALSTM} are less noticeable since outstanding results can be achieved even without augmented data, but still, the former reaches the highest recall and \gls{Bi-ALSTM} is the most robust in terms of overall performance. 

Observing confusion matrices in the second column in Figure~\ref{fig:confuxion}, all models reveal roughly the same pattern, as opposed to the corresponding results on the first. This confirms that \emph{augmentation encourages models to learn associating timing info rather than payload features} in the classification task. 

\subsection{Impact of Feature Arrangement}

\begin{figure}[t]
    \centering
    \includegraphics[width=0.8\linewidth]{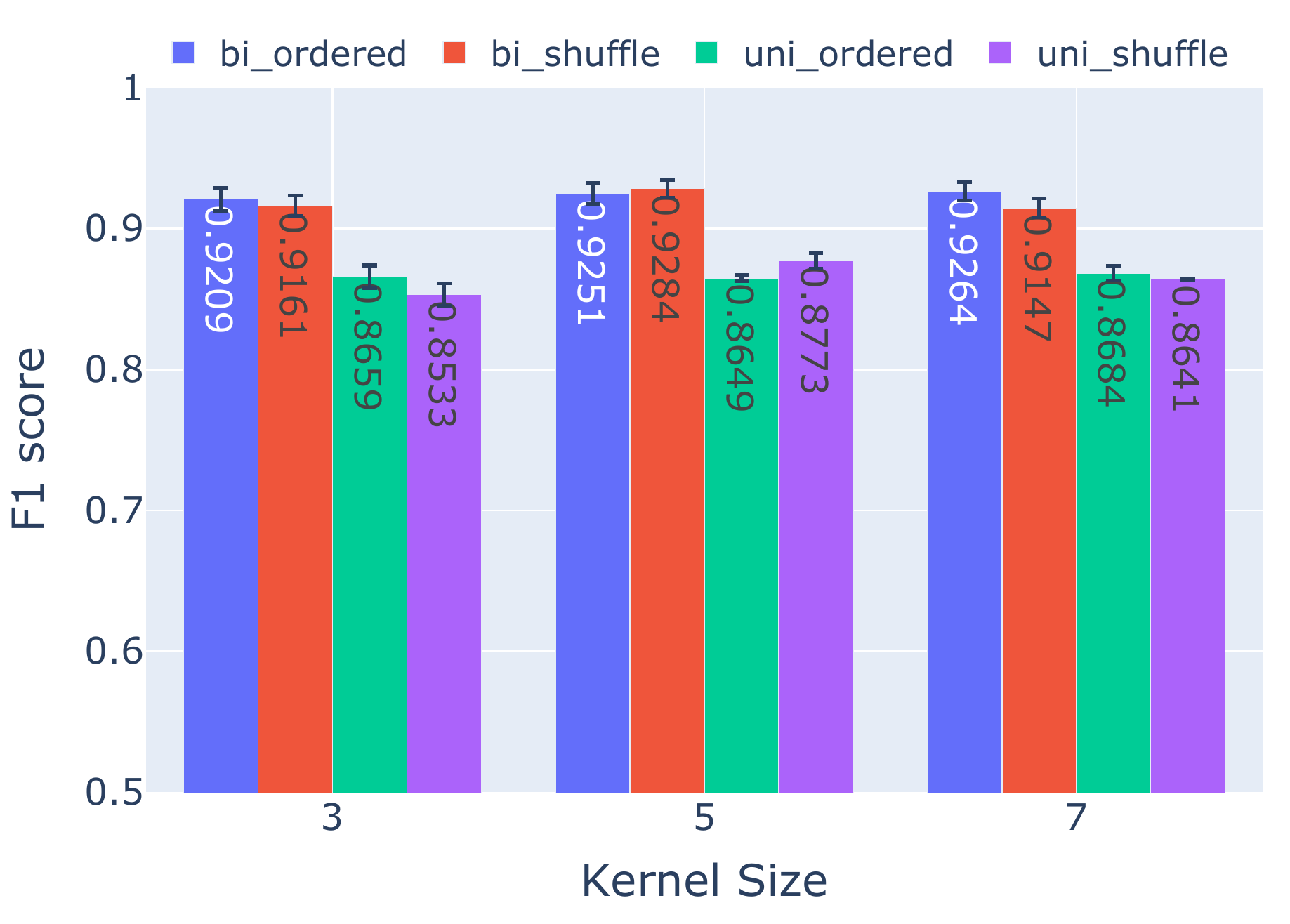}
    \vspace*{-1em}
    \caption{The F1 scores attained by (Bi-)\gls{ConvLSTM} with different kernel sizes and different order of features.}
    \label{fig:conv}
\end{figure}

We investigate the influence of the feature arrangement and  kernel size on the performance of \gls{ConvLSTM}. For this, we experiment with 3 different kernel sizes, namely (3, 5, 7), and two sets of feature arrangements. Specifically, the 1D feature vectors are logically ordered and randomly shuffled. Note that most of the features listed in Table~\ref{table:features} in the Appendix are computed for forward traffic only, backward traffic only, and bidirectionally. Logically ordered features means that they follow an alternating order of forward, backward, and bi-direction. In each experiment, we train both unidirectional \gls{ConvLSTM} and Bi-\gls{ConvLSTM} with the augmented dataset for 10 epochs and repeat the process 5 times. The mean and error bars of the F1 score on the cross-evaluation dataset (CIC-IDS-2017) are illustrated in Figure~\ref{fig:conv}.

Intuitively, one might expect \gls{ConvLSTM} would only work with sequential data possessing clear spatial information, such as videos. However, we find that \gls{ConvLSTM} is robust to 1D traffic features regardless of their arrangement. Indeed, the results in Figure~\ref{fig:conv} demonstrate that there is no significant gap between the model trained with logically ordered features or randomly shuffled ones. For most cases, the mean in the former case is slightly higher than in the latter, while the error bars show a large degree of overlap. 

\label{sec:featurearrange}

We also find that although F1 scores tend to rise sightly with larger kernel sizes when the \gls{ConvLSTM} is trained with the ordered feature arrangement, this increase is not significant. Considering the growth in computation overhead with using a larger kernel size, we argue that training both Bi-\gls{ConvLSTM} and \gls{Bi-ALSTM} with a kernel size equal to 3 (which was also the case for the results presented in Table \ref{table:overall_results}) is sufficient.

\subsection{Performance Gains of \gls{Bi-ALSTM}}
\begin{figure}[h!]
    \centering
    \includegraphics[trim=0 0 0 25, clip, width=\linewidth]{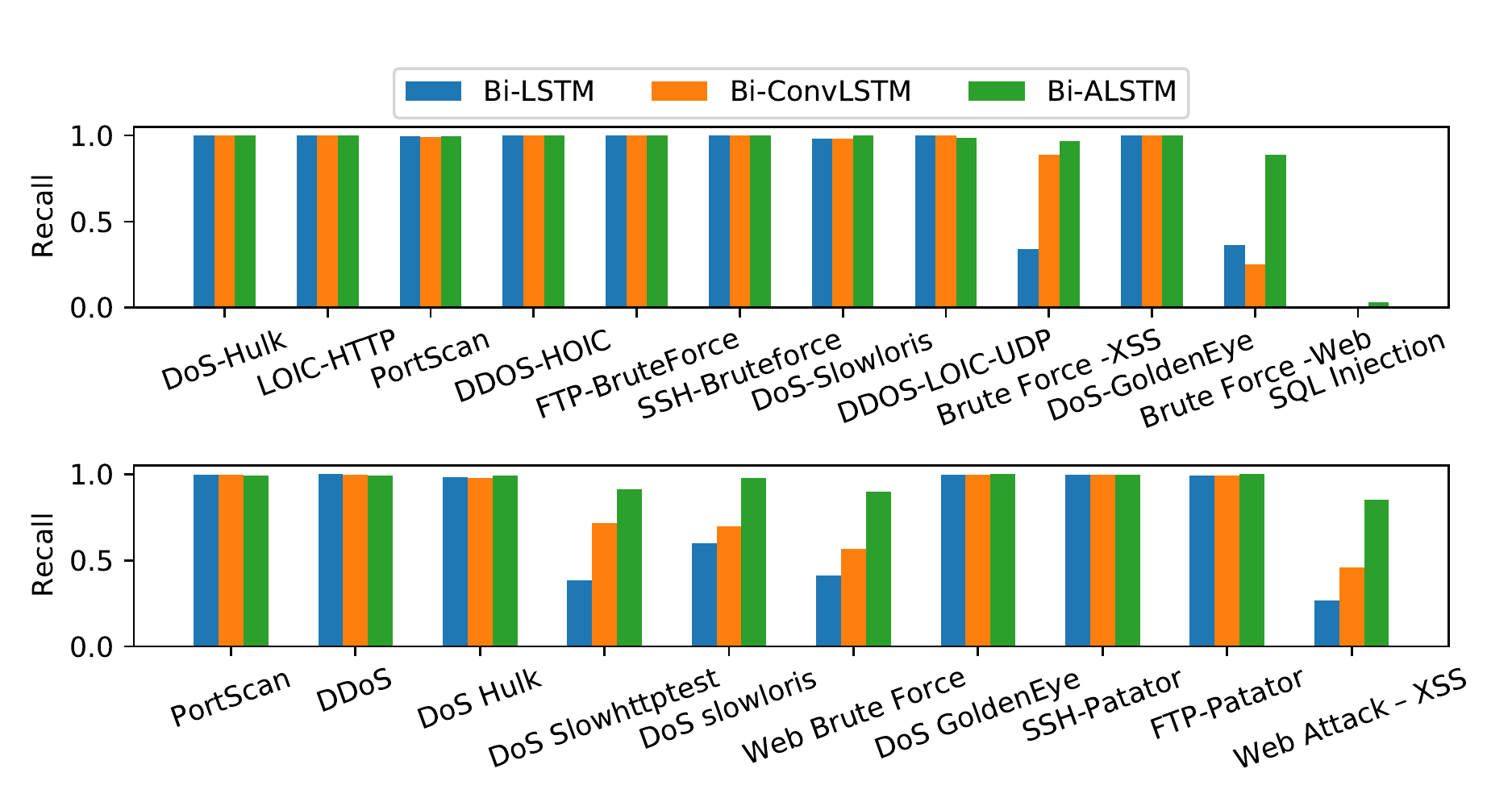}
    \vspace*{-1em}
    \caption{Detection rate (recall) of each type of traffic evaluated on CSE-CIC-IDS2018 (top) and CIC-IDS-2017 (bottom).}
    \label{fig:recall_by_type}
\end{figure}

\gls{Bi-ALSTM} not only yields the highest overall detection rate (recall), but also reliably detects each type of cyber attacks, as illustrated in Figure \ref{fig:recall_by_type}. Both \gls{Bi-LSTM} and Bi-ConvLSTM have difficulty recognizing web bruteforcing, \gls{XSS}, and Slowloris attacks, whereas \emph{\gls{Bi-ALSTM} attains up to 3$\times$ higher detection rates}. The only exception is SQL injection, which cannot be detected by all the algorithms. This is because there are only 53 instances of this attack, merely accounting for 0.0006\% of the entire dataset, which is insufficient for the model to learn a reliable decision boundary for classification. 

\begin{figure}[h]
    \centering
    \includegraphics[width=\columnwidth]{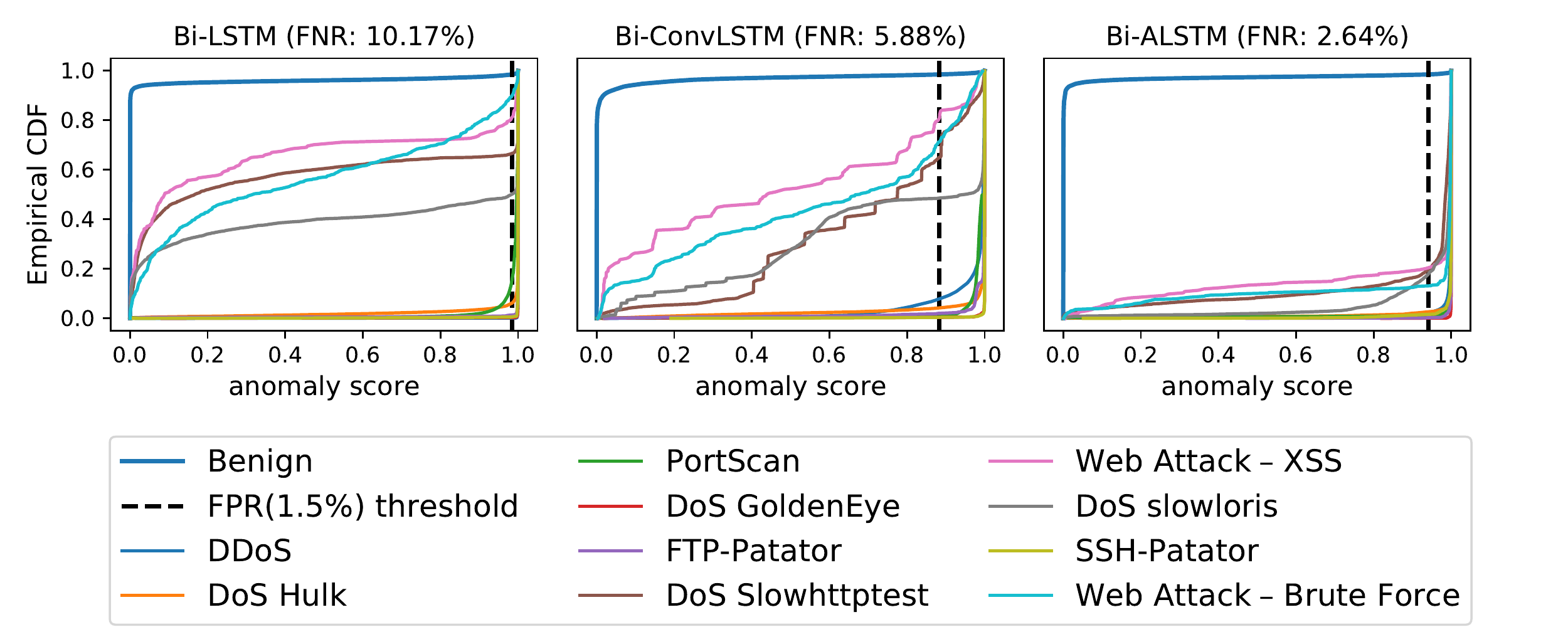}
    \vspace*{-1em}
    \caption{ECDF of the anomaly scores with respect to each type of traffic in CIC-IDS-2017 given by \gls{Bi-LSTM}, Bi-ConvLSTM and \gls{Bi-ALSTM}. All models trained with augmented dataset.}
    \label{fig:ecdf}
\end{figure}

We evaluate the quality of the anomaly scores approximated by Bi-(Conv)LSTM and \gls{Bi-ALSTM}. The anomaly score is the value output by the model. Since the activation function of the last layer is softmax, the output is squeezed between $[0, 1]$ and the higher the value, the more anomalous a flow is regarded. System administrators routinely customize an anomaly threshold to lower the \gls{FPR}. Figure \ref{fig:ecdf} plots the \gls{ECDF} of each type of traffic in CIC-IDS-2017 given by the three algorithms, in which the blue line corresponds to benign traffic. The black dashed line is the threshold that sets the \gls{FPR} to 1.5\%, and the area under the other lines to the left of the threshold line represents the proportion of attacks that would be misclassified. We find that \gls{Bi-ALSTM} delivers the lowest \gls{FNR} (2.63\%) compared with Bi-LSTM (10.17\%) and Bi-ConvLSTM (5.87\%).

\subsection{Computational Overhead}
While \name is primarily designed as an offline \gls{NIDS}, employing it for online NID is also feasible. Table~\ref{table:computeMAC} details the \gls{MACs} the benchmark models and our Bi-ConvLSTM/-ALSTM structures require for a single traffic flow inference, as well as their number of parameters. (CNN-)\gls{Bi-LSTM} are the most computationally expensive, given that multiple fully-connected layers are embedded in the LSTM unit. In contrast, \emph{Bi-ConvLSTM/-ALSTM are relatively lightweight, both involving fewer computations and parameters}. Deploying \name as an online system next to routers or organizational gateways equipped with a GPU or TPU should thus be straightforward.

Given that edge AI platforms are now available, e.g., Nvidia Jetson Nano \cite{jetson}, running \name on constrained small-business/ home routers is within reach. Results in Table \ref{table:computeMAC} reveal that Bi-ConvLSTM/-ALSTM can handle 4.5/3.5 Mflows per second, which confirms our practicality assessment.

\begin{table}[t]
\small
\centering
\bgroup
\def\arraystretch{1.2}
\setlength{\tabcolsep}{5pt}
\begin{tabular}{c|ccc}
\Xhline{2\arrayrulewidth}
Model & \gls{MACs}(k) & Parameters(k) & \begin{tabular}[c]{@{}l@{}}Edge GPU\\(Mflow/s)\end{tabular}\\ \hline
MLP & 5.7 & 5.8 & 41.4\\
CNN & 3.2 & 1.5 & 73.7\\
Autoencoder & 10.3 & 10.6 & 22.9\\
OC-NN & 5.2 & 5.2 & 45.4\\
Kitsune & 0.7 & 0.8 & 337.1\\
DAGMM & 5.3 & 5.4 & 44.5\\
Bi-LSTM & 102.2 & 100.7 & 2.3\\
CNN-Bi-LSTM & 116.8 & 112.7 & 2.0\\
Bi-ConvLSTM & 51.8 & 2.7 & 4.5\\
Bi-ALSTM & 66.8 & 41.4 & 3.5\\ 
\Xhline{2\arrayrulewidth}
\end{tabular}
\egroup
\caption{Computation overhead in terms of \gls{MACs} per inference instance and number of parameters for each model. The last column presents the number of flows (in millions) an edge GPU (Jetson Nano) can process per second.}
\label{table:computeMAC}
\end{table}

Another key merit of \name is that the system inherently analyses consecutive traffic between pairs of hosts, which is easy to integrate into an \gls{IPS}, without the need for collecting statistics of potentially malicious hosts until reaching full confidence about decisions to enforce. Recall that our system directly gives prediction results about the traffic flows generated between two hosts during a short interval, offering comprehensive contexts to the \gls{IPS} with low \gls{FP} risks. Dynamic firewall rules can also be updated effortlessly, since the atomic processing input of \name originates from the same pair of hosts.

\section{Discussion}
\label{sec:evasion}
Lastly, we discuss the robustness of our system against different evasion attacks.

\textbf{IP Spoofing:} 
IP addresses can be spoofed with little effort, which is also a common approach to generating \gls{DDoS} attacks. Flow-based \gls{NIDS} may be ineffective in preventing such traffic because `identities' are changed frequently. However, it is worth noting that IP spoofing can only be used to initiate stateless \gls{DDoS}, given that any responses from the victim is not guaranteed to be routed back to the attacker. Existing countermeasures such as TCP half-open and ICMP threshold are capable of mitigating those issues. For application-layer \gls{DDoS}, 
attackers must control the real IP addresses to maintain the states, where \name will not be fooled.

\textbf{{Traffic Encryption:}}
Traffic Encryption was proposed for evasion attacks \cite{stinson2008towards}, whereby malicious payload is hidden in an encrypted channel. This is however only effective against \gls{NIDS} that examine the syntax of network communications, such as BotHunter \cite{gu2007bothunter}. \name is designed to extract and analyze timing- and protocol-based statistics. That said, manipulation of payload contents cannot bypass our design.

\textbf{{Adversarial Perturbations:}}
Adding small perturbations to input data may lead to misclassification by \gls{ML} models \cite{co2019}. 
Nevertheless, the existing adversarial attacks often require access to model gradients, structures, or numerous queries for weight approximation. In reality, a \gls{ML}-based \gls{NIDS} would not disclose the details of its neural model and tolerate countless queries. 
Zhang et al. demonstrate the possibility of attacking \gls{ML}-based intrusion detection algorithms by heuristic-based methods without knowing model's information \cite{zhang:2020}, but it would still take 100$\sim$11,000 queries to generate one adversarial sample. Note that \name is intended for continuous and repetitive network attacks, meaning that similar queries would trigger alarms much earlier than discovering a valid adversarial sample. 

Adversarial perturbations are likely to modify every individual feature to create malicious samples, which is not always practical in the networking domain, since the modified flows are not guaranteed to stem from any real traffic. Consider instead a more pragmatic attack scenario where the adversary slows the attack speed by increasing the time between the packets sent. To evaluate the potential impact of this adaptive attack on \name, we first pre-process both CSE-CIC-IDS2018 and CIC-IDS-2017 datasets as follows: (1) we group the packets belonging to malicious activity in the {\ttfamily pcap} traces into flows; (2) in each flow, we alter the timestamps of the attacker's packets by expanding the time gap between the previously received/sent packet and the current one, by a fixed multiplier $ m \in \{1, 2, 4, 8\}$ (packets are not delayed if $m=1$); and (3) we alter the timestamps of the victim's packets to ensure the time gaps between these and the attacker's packets still match those in the original flows. 

PortScan attacks are excluded from both datasets because the majority only consists of 1--2 packets, and applying the logic above will not change their timestamps at all. 
We choose a set of multipliers $ m \in \{1, 2, 4, 8\}$, where the attacker's packets are not delayed if $m=1$. As such, 
We obtain three altered versions of the CSE-CIC-IDS2018 and CIC-IDS-2017 dataset. Each variant of CSE-CIC-IDS2018 is split into a training set (70\% of samples) and a test set (30\%), and we augment all the training sets as detailed in \S\ref{sec:aug}, then retrain the Bi-ALSTM. The altered CIC-IDS-2017 datasets are used for cross evaluation. 
We measure the Percentage Error (PE) with respect to the F1 score, to understand to what extent the model would degrade when facing malicious traffic that is purposely slowed down by different factors, to attempt evasion. Formally, PE wrt. F1 score is defined as:
\[
PE^{F1}_{i, j} = \frac{F1_{i, j} - F1_{i, i}}{F1_{i, i}} \times 100\%, 
\]
where the first subscript denotes the multiplier $m=i$ applied in the dataset used for model training, and the second subscript to slow-down factor in the set used for testing. As shown in Table \ref{table:slowdown}, we find that the maximum PE on the CSE-CIC-IDS2018 is never above 0.35\% and the maximum PE on CIC-IDS-2017 is below 0.58\%. This demonstrates that \emph{manipulating the attack timing has no effective impact on the detection performance of the proposed \name.}

\begin{table}[t]

\small
\centering
\bgroup
\def\arraystretch{1.0}
\begin{tabular}{c|c|cccc}
\Xhline{2\arrayrulewidth}
 &  & \multicolumn{4}{c}{Test (CSE-CIC-IDS-2018)} \\ \cline{2-6} 
 & m & 1 & 2 & 4 & 8 \\ \hline
\multirow{4}{*}{\begin{tabular}[c]{@{}c@{}}Train\\ (IDS-2018)\end{tabular}} & 1 & 0 & -0.08\% & -0.02\% & -0.02\% \\
 & 2 & -0.32\% & 0 & -0.16\% & -0.24\% \\
 & 4 & -0.35\% & 0 & 0 & -0.01\% \\
 & 8 & -0.09\% & +0.2\% & 0 & 0 \\ \Xhline{2\arrayrulewidth}
 &  & \multicolumn{4}{c}{Cross Test (CIC-IDS-2017)} \\ \cline{2-6} 
 & m & 1 & 2 & 4 & 8 \\ \hline
\multirow{4}{*}{\begin{tabular}[c]{@{}c@{}}Train\\ (IDS-2018)\end{tabular}} & 1 & 0 & -0.05\% & -0.11\% & -0.58\% \\
 & 2 & +0.1\% & 0 & +0.2\% & -0.39\% \\
 & 4 & 0 & -0.01\% & 0 & -0.35\% \\
 & 8 & -0.49\% & -0.56\% & -0.11\% & 0 \\
\Xhline{2\arrayrulewidth}
\end{tabular}
\egroup
\caption{Percentage Error wrt. F1 score. Attacker's packets slowed down by factors $\{1, 2, 4, 8 \}$. $m=1$ for original timing. }
\label{table:slowdown}
\end{table}

\section{Related Work}
Network intrusion detection has been the focus of extensive research in the security community. In what follows, we briefly discuss the most relevant work related to ours, highlighting limitations of prior approaches and similarities with the proposed \name, where appropriate.

\textbf{Defenses through Offensive Footprint Profiling.} Modeling the unique malicious nature of network anomalies is effective for detection. BotHunter \cite{gu2007bothunter} builds infection dialogues to describe the dynamic process of Botnet infection, then employs modularized detection engines to identify the footprint of each stage in an attack. BotSniffer \cite{gu2008botsniffer} identifies bot activity by highlighting the spatiotemporal correlations of Command and Control (C\&C) traffic originating from pre-programmed behaviors. 
Profiling malicious code execution paths plays an important role in detecting malware \cite{kolbitsch2009effective, naderi2019malmax}. Likewise, stealth DDoS amplification can be fingerprinted by 
its unique two-stage behavior (i.e., scan and attack) \cite{krupp2016identifying}. These contributions demonstrate that modeling the potential links between different attack phases has merit in practice. 
Unlike previous works, here we reveal how \emph{different stages are common between different large-scale attacks} and why breaking their sequence is essential to thwarting intrusions.

\textbf{Time-invariant \gls{ML} for \gls{NID}.} 
\gls{DL}-based \glspl{NIDS} learn illicit traffic patterns through a spectrum of algorithms, replacing the explicit attack modeling methodology introduced previously. In detecting anomalies, such algorithms
largely performing analysis on a per-sample basis, i.e., using statistical features of a traffic flow, to determine its nature, rather than exploring any potential correlations in network traffic. 
\textit{Supervised Learning} approaches, including RIPPER \cite{lee1998data}, \gls{SVM} \cite{yi2011incremental}, 
and Random Forest \cite{sangkatsanee2011practical}, 
treat anomaly detection as a classification problem, seeking a decision boundary between benign and malicious traffic. \textit{Semi-supervised Learning} methods 
discard anomalous samples during training, and only learn patterns of benign traffic. Kitsune \cite{mirsky2018kitsune} learn to reconstruct benign data via encoder ($\mathcal{E}$) and  decoder ($\mathcal{D}$) networks. Samples with high reconstruction errors, i.e., $||x - \mathcal{D}(\mathcal{E}(x)) ||_{2}$, are deemed as malicious. One-Class Deep SVDD \cite{ruff2018deep} believes benign samples can be enclosed by a hyper-sphere, whereas anomalous ones are distinct from the center. Thus, Deep SVDD learns a non-linear transformation that maps innocuous samples into a feature space where the majority of them can be surrounded by a small hyper-sphere.  Statistical approaches assume that benign data in nature are densely distributed in the feature space, while anomalies (outliers) are scattered. Dense areas can be approximated by Deep Gaussian Mixture models \cite{zong2018deep} or Generative Adversarial Networks \cite{li2019mad}.

\textbf{Time-sensitive \gls{ML} for \gls{NID}} relies on temporal context along with a sample, to detect any intrusion. 
\gls{NIDS} that employ this approach are sparse. 
More commonly, it is \gls{HIDS} \cite{du2019lifelong, shen2018tiresias, su2019robust, liu2020deep} that utilize time-sensitive models, such as \gls{LSTM} and \gls{RNN}, because the target data (system calls, logs and security events) present obvious semantic meaning and potential temporal dependencies. Attention-based Graph Neural Networks \cite{deng2021graph} can also be used to model high-dimensional time-series data and spot anomalies. Alternatively, USAD \cite{audibert2020usad} handles time-sensitive tasks by segmenting time-series data into fixed-size windows, and uses adversely trained autoencoders to detect intrusions or anomalies.
Recent studies attempt to model temporal correlations within network attacks and propose a range of RNN-based algorithms \cite{diro2018leveraging, li2018semantic}. However, an appropriate threat model detailing what temporal information is relevant to \gls{NID} is missing. 
Moreover, the training inputs are often randomly sampled, which suppresses relevant temporal information and makes \gls{NID} effectiveness questionable. Our \name design sets to address this particular issue and \emph{takes a dynamic view to cyber attacks, so as to identify possible temporal relationships that exist among different types of attacks}, thereby building a well-directed defensive approach.  

\section{Future Work}
As we discuss in Section \ref{sec:evasion}, existing adversarial attacks on \gls{NIDS} add perturbations to the statistical features of traffic flows. There is no guarantee that perturbed features can be mapped back to a sequence of packets to be transmitted in practice. It remains unclear whether conducting adversarial attacks by directly shaping consecutive packet sizes and inter-arrival times can deceive ML-based NIDS. We deem this topic as important, since it could further shed light on the robustness and reliability of our method. 

On the other hand, \gls{Bi-ALSTM} is a supervised algorithm that demands a significant amount of data for training, but acquiring up-to-date datasets is not always feasible, given the stealth nature of cyber attacks. Unfortunately, existing semi-supervised algorithms still focus on per-flow classification and neglect temporal context, resulting in the undesirable performance seen in Table \ref{table:overall_results} (Autoencoder, OC-NN, Kitsune, DAGMM). Instead of approximating the distribution of benign flows, estimating the stochastic process of consecutive benign traffic may provide higher reliability, which is also an interesting topic for future study.

\section{Conclusions}
In this paper, we show that large-scale network threats with potential high-impact can be tackled in their early stages, if correctly recognizing the unique temporal dependencies of malicious flows, and we propose \name to effectively detect such incipient attacks. \name incorporates a novel data augmentation technique to enhance the generalization ability of supervised algorithms and we design an ensemble \gls{Bi-ALSTM} as the core intrusion detection logic. Extensive results demonstrate that our ensemble  structure outperforms a wide range of benchmarks, attaining up to 3$\times$ higher detection rates, under different network environments. Finally, we discuss computation overhead and robustness to evasion attacks, making the case for the feasibility of deploying  \name alongside threat prevention logic in real-world settings. 

\section*{Acknowledgments}
This material is based upon work supported
by Arm Ltd and Scotland's Innovation Centre for sensing, imaging and Internet of Things technologies (CENSIS).

\bibliographystyle{ieeetr}
\bibliography{main}

\end{document}